\pdfoutput=1
\documentclass{article}

\usepackage{graphicx,epsfig,color}
\usepackage{cite}
\usepackage[english]{babel}
\usepackage{amssymb,amsfonts,amsmath,bm,bbm}
\usepackage{verbatim}
\usepackage{tikz}
\usepackage{bbold}
\usepackage{hyperref}

\newcommand{\vev}[1]{{\left\langle\, #1\, \right\rangle}}
\def\tT{\widetilde{T}}
\def\cL{{\cal L}}

\def\f{{f_\pi}}

\newcommand{\mt}[1]{\textrm{\scriptsize #1}}
\def\Nf{N_\mt{f}}
\def\Nc{N_\mt{c}}

\newcommand{\be}{\begin{equation}} \newcommand{\ee}{\end{equation}}
\newcommand{\bea}{\begin{eqnarray}} \newcommand{\eea}{\end{eqnarray}}

\usepackage[top=2.8cm, bottom=2.8cm, left=2.5cm, right=2.5cm]{geometry}	% for manually adjusting margins
\numberwithin{equation}{section}

\begin{document}

\begin{flushright}
HIP-2022-4/TH
\end{flushright}

\begin{center}

\centering{\Large {\bf Revisiting the chiral effective action in holographic models}}

\vspace{8mm}

\renewcommand\thefootnote{\mbox{$\fnsymbol{footnote}$}}
Carlos Hoyos,${}^{1,2}$\footnote{hoyoscarlos@uniovi.es}
Niko Jokela,${}^{3,4}$\footnote{niko.jokela@helsinki.fi} and
Daniel Logares${}^{1,2}$\footnote{dani.logares@gmail.com} 

\vspace{4mm}
${}^1${\small \sl Department of Physics} \\
{\small \sl Universidad de Oviedo}, {\small \sl c/ Federico Garc\'{\i}a Lorca 18, ES-33007 Oviedo, Spain} 

\vspace{2mm}
\vskip 0.2cm
${}^2${\small \sl Instituto Universitario de Ciencias y Tecnolog\'{\i}as Espaciales de Asturias (ICTEA)}\\
{\small \sl  Calle de la Independencia, 13, 33004 Oviedo, Spain}

\vspace{2mm}
\vskip 0.2cm
${}^3${\small \sl Department of Physics} and ${}^4${\small \sl Helsinki Institute of Physics} \\
{\small \sl P.O.Box 64} \\
{\small \sl FIN-00014 University of Helsinki, Finland} 

\end{center}

\vspace{0mm}

\renewcommand\thefootnote{\mbox{\arabic{footnote}}}

\begin{abstract}
\noindent  
We obtain the pion decay constant and coefficients of fourth derivative terms in the chiral Lagrangian for massless quarks in the Witten-Sakai-Sugimoto model. We extract these quantities from the two-pion scattering amplitude, which we compute directly in the holographic dual through tree-level Witten diagrams. Identification of the low energy coefficients in the chiral action is subtle as their values will be shifted when the tower of massive vector bosons are integrated out. Indeed, by a direct comparison with the existing standard procedure of constructing the chiral action with radial modes in the gravity dual, we explicitly show that there are finite 't Hooft coupling corrections that have been missed. This suggests that past derivations of effective actions from holographic models may have to be revisited and future derivations more carefully considered.
\end{abstract}

\newpage
\tableofcontents
\newpage

%%%%%%%%%%%%%%%%%%%%%%%%%%%
%%%%%%%%%%%%%%%%%%%%%%%%%%%
\section{Introduction}
%%%%%%%%%%%%%%%%%%%%%%%%%%%
%%%%%%%%%%%%%%%%%%%%%%%%%%%

Since the inception of the gauge/gravity, or holographic, duality there has been a conscious effort to perceive Quantum Chromodynamics (QCD) in the non-perturbative regime \cite{Witten:1998zw}. We do not know if the holographic dual of QCD exists and, even if it is the case, there is likely no weakly coupled gravity description. To date, quantitative predictions have usually been based on a phenomenological approach where weakly coupled holographic models are fitted to known QCD data obtained through experiments or other non-perturbative approaches. The most developed model in this respect is presumably V-QCD \cite{Jarvinen:2011qe}.

The observables that are relevant in the construction of a holographic model are low energy coupling constants (LECs). These constants enter in the effective QCD action and determine the interactions among hadrons and, most importantly, their values can be inferred from experiments. Chiral perturbation theory (ChPT) (see \cite{Ecker:1994gg,Epelbaum:2008ga,Machleidt:2011zz} for reviews) provides a systematic approach to characterize the LECs based on the approximate flavor symmetry of the microscopic QCD Lagrangian and the spontaneous breaking of this symmetry by a chiral condensate in the QCD vacuum. A holographic model that aspires to quantitatively counterfeit QCD observables in the confined phase should therefore be able to reproduce the chiral effective action with values of the LECs that match the experimental observations.

In this program being able to extract the LECs from the holographic model is fundamental. An early proposal on how to construct the effective action was within the Witten-Sakai-Sugimoto (WSS) model \cite{Witten:1998zw,Sakai:2004cn,Sakai:2005yt} by unveiling the action in the gravity side for modes of the fields dual to mesons. Later works in other models followed a similar approach \cite{DaRold:2005mxj,Hirn:2005nr,DaRold:2005vr,Erlich:2005qh,Harada:2006di,Chivukula:2006fxj,Panico:2007qd,Domenech:2010aq,Harada:2010cn,Colangelo:2012ipa,Domokos:2014ura,Harada:2014lza,Espriu:2020ise,Liu:2022out,Lyubovitskij:2022rod}. In this work we follow a different path, we extract the LECs from the low-energy scattering amplitude of pions, that we compute directly from Witten diagrams in the gravity dual following the method developed in \cite{Hoyos:2019kzt,Hoyos:2020fjx}. To be definite, we focus on the WSS model with two flavors of massless quarks and find that the coefficients originating from pion self-interactions in the gravity dual were misidentified.

\medskip

\paragraph{Summary of the discrepancy with existing literature}	
	
Let us exbound the nature of the discrepancy, by examining the original result of Sakai and Sugimoto in \cite{Sakai:2005yt}. We postpone a detailed description of the chiral Lagrangian and scattering amplitudes in section~\ref{sec:chiralEFT} to keep the discussion here more concise.

For $\Nf$ flavors, the pion field is a $\Nf\times \Nf$ unitary matrix $U=e^{2i\Pi/\f}$, with $\Pi$ a Hermitean matrix. In addition, there is a tower of massive vector meson fields  $v_\mu^n$, $n=1,2,3,\ldots$. Higher $n$ corresponds to higher mass. From the dual gravity point of view, $n$ labels Kaluza-Klein modes in the holographic radial direction. The effective action (in the absence of sources) was found to be of the form
\begin{equation}\label{eq:actionWSS}
\begin{split}
\cL=&-\text{tr}\left(\partial_\mu\Pi\partial^\mu \Pi \right)-\frac{1}{3 \f^2} \text{tr}\left[ \Pi, \partial_\mu \Pi\right]^2+\frac{1}{2 e_S^2\f^4}\text{tr}\left[ \partial_\mu \Pi,\partial_\nu\Pi\right]^2\\
&+\sum_n 2\text{tr}\left(\partial_{[\mu} v_{\nu]}\right)^2+m_{v^n}^2(v_\mu^n)^2 + \frac{2b_{v^n\pi\pi}}{\f^2}\text{tr}\left(\partial_{[\mu}v_{\nu]}\left[\partial^\mu \Pi,\partial^\nu \Pi\right] \right)\ , 
\end{split}
\end{equation}
where $e_S^2\f^2\simeq 0.51$. Focusing on the quartic terms in the pion field $O(\partial^2\Pi^4)$, the na\"{\i}ve LECs in this action\footnote{Their precise definition can be found in Eqs.~\eqref{eq:pionaction} and \eqref{eq:l3action}.} would be the same as for the Skyrme model \cite{Skyrme:1961vq,Skyrme:1961vr,Skyrme:1962vh}. For $\Nf=3$ the LECs satisfy the relations
\begin{equation}\label{eq:WSSLECs}
L_2^{SU(3)}=2 L_1^{SU(3)} \ , \ L_3^{SU(3)}=-3L_2^{SU(3)},\ \ L_1^{SU(3)}=\frac{1}{32 e_S^2}  \ .
\end{equation}
The first relation actually always holds in the large-$\Nc$ limit, so there are only two independent coefficients (this is true for any value of $\Nf$). For $SU(2)$ the term proportional to $L_3$ can be recast as the term proportional to $L_1$, so that for the WSS model
\begin{equation}
L_1^{SU(2)}=L_1^{SU(3)}+\frac{1}{2}L_3^{SU(3)}=-L_2^{SU(3)}=-L_2^{SU(2)}\,.
\end{equation}
However, the action \eqref{eq:actionWSS} is not in the standard form, in particular, the last term capturing the coupling between the vector mesons and the pions is not the expected one from hidden local symmetry (HLS) considerations. In order to correct for this, one redefines the vector meson fields as follows
\begin{equation}\label{eq:mesonshift}
\hat{v}^n_\mu=v_\mu+\frac{b_{v^n\pi\pi}}{2\f^2}\left[ \Pi,\partial_\mu\Pi\right]\,.
\end{equation}
After this redefinition the action becomes
\begin{equation}\label{eq:actionWSS2}
\cL=-\text{tr}\left(\partial_\mu\Pi\partial^\mu\Pi\right)+\sum_n 2\text{tr}\left(\partial_{[\mu} \hat{v}_{\nu]}\right)^2+m_{\hat{v}^n}^2(v_\mu^n)^2 -2 g_{v^n\pi\pi}\text{tr}\left(\hat{v}_{\mu}^n\left[\Pi,\partial^\mu \Pi\right] \right)\,.
\end{equation}
So, according to this, the actual prediction before vector mesons have been integrated out is that the LECs are zero. Sakai and Sugimoto computed the pion scattering amplitude to lowest order in momentum (section 3.7 of \cite{Sakai:2005yt}), and found that even though contributions from quartic contact terms cancel out, the contribution from vector meson exchange produces the right amplitude (proportional to $1/\f^2$) as expected from low energy theorems, but a complete calculation showing that the LECs are indeed given by \eqref{eq:WSSLECs} was missing.

This is not the whole story, however, because there are other possible contributions to the LECs in the WSS model that have been neglected so far. They originate from higher derivative corrections in the holographic model and would be relatively suppressed by powers of the 't Hooft coupling. We will compute them and show that these contributions are indeed non-vanishing.

To derive the actual chiral Lagrangian below the rho meson mass one can integrate out the massive vector mesons to produce an effective action that contains only the pion. In this case the LECs are generally shifted relative to the explicit terms appearing in the actions above. They can be read off from the pion scattering amplitude expanded to the suitable order in momentum, taking into account vector meson exchange contributions. As we will see, the LECs derived in this way coincide with the values quoted by Sakai and Sugimoto, up to the corrections we just mentioned. This coincidence, at least regarding the relative values of the LECs, stems from vector meson dominance, which originates from the meson couplings in \eqref{eq:actionWSS2}. The LECs in \eqref{eq:actionWSS} in this case give the right value because the coupling between the vector mesons and the pions is such that it turns out not to modify their value when the vector mesons are integrated out.

In conclusion, one should be careful when identifying the LECs from an action derived holographically, at least when the action includes massive mesons coupled to the pions, as the low energy pion scattering amplitude can receive meson exchange contributions.  
The actions \eqref{eq:actionWSS} and \eqref{eq:actionWSS2} are related by a field redefinition, so they produce the same low energy scattering amplitude and, when the massive vector mesons are integrated out the same low energy effective action for the pions. The holographic calculation of the scattering amplitude seems to be in accord with the action \eqref{eq:actionWSS2}, the LECs originally quoted by Sakai and Sugimoto originate from exchange diagrams while the new corrections we have found originate from contact interactions.

\vspace{1cm}

The structure of our paper is as follows. In section \ref{sec:chiralEFT} we review the chiral effective action and the ingredients needed to compute the pion scattering amplitude. We also explain the contribution from vector meson exchange and how the values of the LECs are shifted. In section \ref{sec:WSS} we review the salient details of the WSS model and steer focus on the original calculation of the chiral effective action. In section \ref{sec:scattering} we derive the scattering amplitude of pions in the WSS model and extract the LECs. Finally, we contrast our results with those in the existing literature with some further implicatory remarks in section \ref{sec:discussion}. We also generalize our results in the large-$\Nc$ 't Hooft limit to an arbitrary (but fixed) number of flavors. Technical details of the holographic calculation of the scattering amplitude have been compiled in appendix \ref{app:calculation}.

%%%%%%%%%%%%%%%%%%%%%%%%%%%%%%%%%%%%%%%%%%%%%
%%%%%%%%%%%%%%%%%%%%%%%%%%%%%%%%%%%%%%%%%%%%%
\section{Chiral Effective Action}\label{sec:chiralEFT}
%%%%%%%%%%%%%%%%%%%%%%%%%%%%%%%%%%%%%%%%%%%%%
%%%%%%%%%%%%%%%%%%%%%%%%%%%%%%%%%%%%%%%%%%%%%

In this section we will spell out some basics involving the chiral effective Lagrangian. Here and in the following we will have in mind a $SU(\Nc)$ Yang-Mills theory with additional matter, and work in the large-$\Nc$ limit. A theory with two flavors of massless quarks enjoys a $U(2)_R\times U(2)_L$ global symmetry. The diagonal component is the vector symmetry $U(2)_V$ and the off-diagonal part is the axial symmetry $U(2)_A$. The latter is anomalous, though the anomaly vanishes for the non-Abelian component. As we will discuss in section~\ref{sec:discussion}, in the large-$\Nc$ limit this is enough to extract the LECs we are interested in for an arbitrary number of flavors.

When a quark condensate is formed, there is a spontaneous breaking of symmetry $U(2)_R\times U(2)_L \longrightarrow U(2)_V$. The Nambu-Goldstone bosons associated to the broken $SU(2)$ axial symmetry are pions and they can be collected in a $SU(2)$ matrix $\Sigma$ that transforms as
\begin{equation}
 \Sigma\longrightarrow U_L\Sigma U_R^\dagger \ .
\end{equation}
The low energy effective theory of the Nambu-Goldstone bosons is captured by the chiral Lagrangian, that admits a systematic expansion in derivatives of $\Sigma$. To fourth order in derivatives one can have the terms
\begin{equation}\label{eq:pionaction}
 \mathcal{L}=-\frac{\f^{2}}{4}\textrm{Tr}\left(\partial_\mu\Sigma^{\dagger}\partial^\mu\Sigma\right)+L_{1}\left(\text{\textrm{Tr}}\left(\partial_{\mu}\Sigma^{\dagger}\partial^{\mu}\Sigma\right)\right)^{2}+L_{2}\left(\text{\textrm{Tr}}\left(\partial_{\mu} \Sigma^{\dagger}\partial_{\nu}\Sigma\right) \right)^2 \ .
\end{equation}
The large-$\Nc$ scaling of the coefficients is $\f^2\sim L_i\sim \Nc$ \cite{Manohar:1998xv}. For three flavors there is an additional term
\begin{equation}\label{eq:l3action}
L_{3}\text{\textrm{Tr}}\left(\left(\partial_{\mu}\Sigma^{\dagger}\partial_{\nu}\Sigma \right)^2\right) \ .
\end{equation}
But for $SU(2)$ there is the following relation
\begin{equation}
 \text{\textrm{Tr}}\left(\left(\partial_{\mu}\Sigma^{\dagger}\partial_{\nu}\Sigma \right)^2\right)=\frac{1}{2}\left(\text{\textrm{Tr}}\left(\partial_{\mu}\Sigma^{\dagger}\partial^{\mu}\Sigma\right) \right)^2 \ .
\end{equation}
Therefore, to compare with other models where the action was written for three flavors one should set $L_1^{SU(2)}=L_1^{SU(3)}+\frac{1}{2}L_3^{SU(3)}$.

The axial current can be obtained by first adding gauge fields for the right-handed and left-handed symmetries by promoting the derivative to a covariant derivative with the left-handed and right-handed gauge fields $L_\mu$ and $R_\mu$
\begin{equation}
\partial_\mu \Sigma\longrightarrow D_\mu \Sigma=\partial_\mu \Sigma+i L_\mu \Sigma-i \Sigma R_\mu \ .
\end{equation}
Considering the $SU(2)$ generators $\tau^a=\frac{1}{2}\sigma^a$, $a=1,2,3$, with $\sigma^a$ the Pauli matrices, the components of the left-handed and right-handed currents are
\begin{equation}
\begin{split}
&J_L^{a\mu}=\frac{\delta \mathcal{L}}{\delta L_\mu^a}=-\frac{i\f^2}{4}\text{\textrm{Tr}}\left(\left(\Sigma\partial^\mu \Sigma^\dagger-\partial^\mu\Sigma\Sigma^\dagger\right) \tau^a\right)+O(\partial^3) \\
&J_R^{a\mu}=\frac{\delta \mathcal{L}}{\delta R_\mu^a}=\frac{i\f^2}{4}\text{\textrm{Tr}}\left(\left(\partial^\mu \Sigma^\dagger\Sigma-\Sigma^\dagger\partial^\mu\Sigma\right) \tau^a\right)+O(\partial^3)  \ . 
\end{split}
\end{equation}
We will introduce the pion field using the exponential parametrization
\begin{equation}
\Sigma=\exp\left(\frac{i}{\f}\bm{\pi}\cdot\bm{\sigma}\right)=\exp\left( \frac{i}{\f}\pi^a\sigma^a\right)\ .
\end{equation}
The axial current in this case is
\begin{equation}\label{eq:J5}
J_5^{a\mu}=J_L^{a\mu} -J_R^{a\mu}=-\f \partial^\mu \pi^a+\frac{2}{3\f}\left((\bm{\pi}\cdot\bm{\pi})\partial^\mu \pi^a-\frac{1}{2}\partial^\mu(\bm{\pi}\cdot\bm{\pi})\pi^a\right)+\ldots \ .
\end{equation}
The Lagrangian density expanded to fourth order in the pion field is
\begin{equation}\label{eq:pionLag}
\begin{split}
\mathcal{L}=&-\frac{1}{2}\partial_\mu\bm{\pi}\cdot \partial^\mu \bm{\pi}-\frac{1}{6\f^2}\left((\partial_\mu\bm{\pi}\cdot \bm{\pi})^2-(\bm{\pi}\cdot \bm{\pi})(\partial_\mu \bm{\pi} \cdot \partial^\mu\bm{\pi}) \right)\\
&+\frac{4L_1}{\f^4}(\partial_\mu \bm{\pi} \cdot \partial^\mu\bm{\pi})^2+\frac{4L_2}{\f^4}(\partial_\mu \bm{\pi} \cdot \partial_\nu\bm{\pi})^2+O(\partial^2 \pi^6) \ .
\end{split}
\end{equation}

%%%%%%%%%%%%%%%%%%%%%%%%%%%%%%%%%%%%%%%%%%%%%
\subsection{Pion scattering amplitude}
%%%%%%%%%%%%%%%%%%%%%%%%%%%%%%%%%%%%%%%%%%%%%

\begin{figure}[thb!]
\begin{center}
\includegraphics[scale=1.]{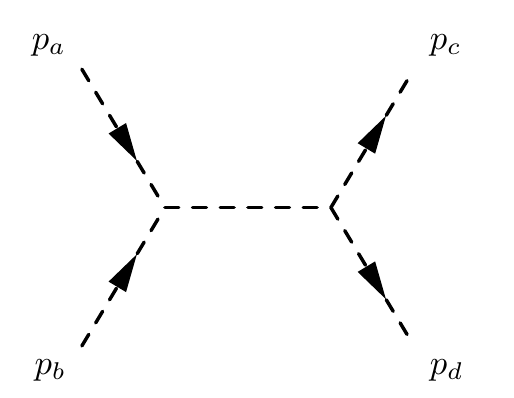}
\includegraphics[scale=1.]{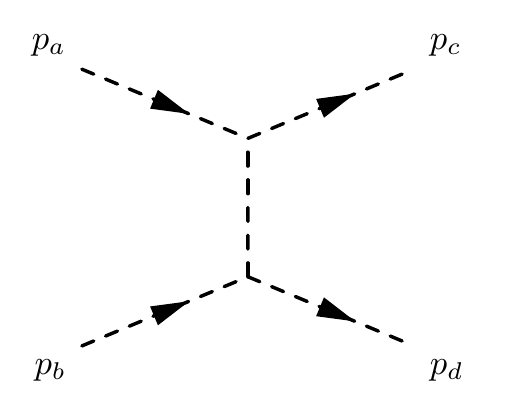}
\includegraphics[scale=1.]{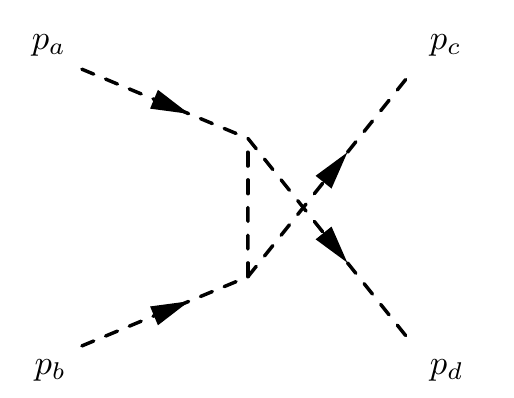}
\caption{The three kinematically non-identical 2-to-2 elastic scattering processes. All propagating particles are pions, hence all legs are represented by dashed lines. Left: $s$-channel, Middle: $t$-channel, and Right: $u$-channel.}\label{fig:stu}
\end{center}
\end{figure}

Let us now discuss how the scattering amplitude can be extracted from the chiral Lagrangian introduced above. The elastic scattering amplitude for two pions
\begin{equation}
 \pi^a(p_a)+\pi^b(p_b)\longrightarrow \pi^c(p_c)+\pi^d(p_d) 
\end{equation}
is given by the T-matrix element 
\begin{equation}
 {\cal T}_{ab,cd}=(2\pi)^4 \delta^{(4)}(p_a+p_b-p_c-p_d){\cal M}_{ab,cd} \ .
\end{equation}
The function ${\cal M}$ is determined by a single scalar function $A(s,t,u)=A(s,u,t)$ defined by the isospin decomposition (see, {\emph{e.g.}}, \cite{Bijnens:1994qh,Ecker:1994gg,Ecker:1996yy})
\begin{equation}\label{eq:pionamplitude}
 {\cal M}_{ab,cd}=\delta_{ab}\delta_{cd}A(s,t,u)+\delta_{ac}\delta_{bd} A(t,s,u)+\delta_{ad}\delta_{bc} A(u,t,s) \ ,
\end{equation}
where $s,t,u$ are the Mandelstam variables
\begin{equation}\label{eq:mandelstam}
 s=-(p_a+p_b)^2 \ ,  \ t=-(p_a-p_c)^2,\ \ u=-(p_a-p_d)^2 \ .
\end{equation}
These variables encode the different scattering processes and correspond to the three channels as depicted in Fig.~\ref{fig:stu}.   At $O(p^4)$, the original derivation by Weinberg produces at tree level \cite{Weinberg:1966kf,Weinberg:1978kz} 
\begin{equation}\label{eq:weinbergA}
 A(s,t,u)=\frac{s}{\f^2}+\frac{8 L_1}{\f^4} s^2+\frac{4L_2}{\f^4}(t^2+u^2) \ .
\end{equation}

For massless pions there are additional logarithmic contributions to the amplitude that are introduced by pion loop corrections. The relevant pion diagram has two quartic vertices with two derivatives, that from the pion Lagrangian \eqref{eq:pionLag} have a large-$\Nc$ scaling $\sim 1/\f^4\sim 1/\Nc^2$. The tree-level contributions on the other hand have a scaling $\sim L_i/\f^4\sim 1/\Nc$. Therefore, in the large-$\Nc$ limit the pion loop contributions are relatively suppressed and are thus not captured by the classical holographic dual calculation. The same statement applies to other meson loop corrections, they are suppressed in the large-$\Nc$ limit. Therefore, our calculation will be limited to tree-level on-shell amplitudes.

In order to compare to gauge/gravity models we will consider the axial current correlators for values of the momenta where the pions are on-shell. We start with the two-point function of the axial current, which will be proportional to the pion propagator\footnote{In principle there are additional terms depending on the pion field in \eqref{eq:J5} that introduce pion loop corrections to the axial current correlator, however, as for the scattering amplitude, these are suppressed in the large-$\Nc$ limit.}
\begin{equation}
\vev{J_5^{a\mu}(-p) J_5^{b\nu}(p)}\underset{p^2\to 0}{\approx} \f^2 p^\mu p^\nu\vev{\pi^a(-p)\pi^b(p)}=-i \f^2\delta^{ab}\frac{ p^\mu p^\nu}{p^2} \ .
\end{equation}
This leads to the Ward identity for current conservation
\begin{equation}\label{eq:wardJ5}
-ip_\mu\vev{J_5^{a\mu}(-p) J_5^{b\nu}(p)}\underset{p^2\to 0}{\approx} -\f^2 p^\nu\delta^{ab} \ .
\end{equation}
However, in the absence of an anomaly and quark masses, the axial current must be conserved and one would have expected this Ward identity to vanish. As we will see, this issue is solved in the holographic model when the two-point function correlator is computed and we find a vanishing value (see (\ref{eq:wardfixed})). The reason is that we are missing a contact term in landing on (\ref{eq:wardJ5}) whose origin could be understood from contributions to the axial current other than the gradient of the pion field.

Similarly, the four-point function of the axial current will be proportional to the one of the pions
\begin{equation}\label{eq:axial4p}
 \vev{J_5^{a\mu_a}(p_a) J_5^{b\mu_b}(p_b)J_5^{c\mu_c}(-p_c)J_5^{d\mu_d}(-p_d)}\underset{p_i^2\to 0}{\approx} \f^4 p_a^{\mu_a}p_b^{\mu_b}p_c^{\mu_c}p_d^{\mu_d}\vev{\pi^{a}(p_a)\pi^{b}(p_b)\pi^{c}(-p_c)\pi^{d}(-p_d)} \ . 
\end{equation}
From the pion correlator we are interested just in the leading pole contribution, which gives us the pion scattering amplitude through the LSZ reduction formula
\begin{equation}\label{eq:pion4p}
 \vev{\pi^a(p_a)\pi^b(p_b)\pi^c(-p_c)\pi^d(-p_d)}_c\underset{p_i^2\to 0}{\approx} (2\pi)^4 \delta^{(4)}(p_a+p_b-p_c-p_d) \frac{i{\cal M}_{ab,cd}}{p_a^2p_b^2p_c^2p_d^2} \ ,
\end{equation}
where ${\cal M}_{ab,cd}$ in this expression is the amplitude \eqref{eq:pionamplitude}.

%%%%%%%%%%%%%%%%%%%%%%%%%%%%%%%%%%%%%%%%%%%%%
\subsection{Vector boson contributions}\label{sec:weinbergact}
%%%%%%%%%%%%%%%%%%%%%%%%%%%%%%%%%%%%%%%%%%%%%

Our discussion thus far has been focusing on the chiral effective action at low energies, where heavier mesons have been integrated out. However, in most holographic examples the effective action includes massive vector bosons coupled to the pions. Whether the vector bosons have been integrated out or not affects the value of the LECs in the pion action. Namely, starting from an action 
\begin{equation}
{\cal L}_{UV}=\mathcal{L}_\pi(\f,L_1,L_2)+\mathcal{L}_{\cal V} \ ,
\end{equation}
where $\mathcal{L}_\pi(\f,L_1,L_2)$ is \eqref{eq:pionaction} with the coefficients $\f,L_1$, and $L_2$ and the last term $\mathcal{L}_{\cal V}$ is the action for the vector bosons. At sufficiently low energies, the vector bosons can be integrated out, resulting in an action that only contains the pions but with modified coefficients
\begin{equation}
{\cal L}_{IR}=\mathcal{L}_\pi(\hat{f}_\pi,\hat{L}_1,\hat{L}_2) \ .
\end{equation}
Let us show this explicitly for the case when the structure of the effective action follows the Hidden Local Symmetry (HLS) approach \cite{Bando:1984ej,Bando:1987br}, which is commonly discussed in the context of holographic models, where it arises from gauge symmetries in the gravity dual. 

In the HLS approach the pion matrix is factorized
\begin{equation}
 \Sigma(x)=\xi_L^\dagger(x) \xi_R(x) \ .
\end{equation}
In addition to the left- and right-moving flavor symmetries there is an emergent gauge symmetry, so that the fields $\xi_{L,R}$ transform as
\begin{equation}
 \xi_L(x)\to h(x) \xi_L(x) U_L^\dagger \ , \ \xi_R(x)\to h(x) \xi_R(x) U_R^\dagger \ . 
\end{equation}
The massive vector boson is identified with the gauge field for the hidden symmetry and the form of the action is constrained by demanding local gauge invariance. The HLS action is directly related to the non-linear realization of local chiral transformations introduced by Weinberg \cite{Weinberg:1968de} by going to the unitary gauge (see, \emph{e.g.}, \cite{Bando:1987br})
\begin{equation}
 \xi_L^\dagger(x)=\xi_R(x)=e^{\frac{i \pi^a(x)}{\f}\frac{\sigma^a}{2}} \ .
\end{equation}
Let us denote $\bm{V}_\mu=V_\mu^a\sigma^a$ as the gauge field associated to the vector boson and $\bm{V}_{\mu\nu}=\partial_\mu \bm{V}_\nu-\partial_\nu \bm{V}_\mu+ g \bm{V}_\mu \times \bm{V}_\nu$ as its field strength with $g$ an effective gauge coupling. In addition to the terms shown in \eqref{eq:pionLag}, the effective action has the vector boson contributions
\begin{equation}
 {\cal L}_V=-\frac{1}{4}\bm{V}_{\mu\nu}\cdot \bm{V}^{\mu\nu}-\frac{1}{2}m_V^2\left( \bm{V}_\mu-\frac{g_{V\pi\pi}}{m_V^2}\bm{\pi}\times \partial_\mu \bm{\pi}\right)^2\ ,
\end{equation}
where $m_V$ is the mass of the vector boson and $g_{V\pi\pi}$ is the coupling to the pion field. The expansion of the invariant mass term for the vector boson includes a cubic coupling with the pions and a $O(\partial^2 \pi^4)$ coupling of the same form as the terms shown in the first line of \eqref{eq:pionLag}.\footnote{Any additional free parameters in the HLS model are absorbed in the coefficients of the gauge-fixed effective action in such a way that the HLS model is equivalent to the non-linear sigma model for tree-level on-shell amplitudes}, see, {\emph{e.g.}}, \cite{Bando:1987br,Harada:2003jx}. 
However, in the pion scattering amplitude this additional vertex contribution is canceled out by the leading contribution from vector boson exchange, so the $O(p^2)$ pion scattering amplitude does not change. In general, there can be vector bosons transforming as adjoint fields of the chiral symmetry with different couplings where this cancelation does not happen, and additional couplings to the vector bosons that appear at $O(p^4)$ and so can also affect the amplitude.

There are, however, $O(p^4)$ contributions that add up to the contributions shown in \eqref{eq:weinbergA}. The tree-level contribution to the pion scattering amplitude from the vector meson exchange has the form \cite{Truong:1987ex,Dobado:1989qm}
\begin{equation}
 A_V(s,t,u)=\frac{g_{V\pi\pi}^2}{m_V^2}\left[\frac{t(s-u)}{m_V^2-t}+\frac{u(s-t)}{m_V^2-u} \right] \ .
\end{equation}
Expanding for small momenta $m_V^4 \gg s^2,t^2,u^2$, and using the fact that here the pions are massless particles yields
\begin{equation}
 s(t+u)=-s^2,\ \ 2 u t=s^2-t^2-u^2 \ .
\end{equation}
Therefore, the vector boson contribution to the low energy scattering amplitude reduces to
\begin{equation}\label{eq:effectiveLEC}
 A_V(s,t,u)=\frac{g_{V\pi\pi}^2}{m_V^4}\left[t^2+u^2-2s^2 \right] \ .
\end{equation}
We can compare this expression with Weinberg's amplitude  \eqref{eq:weinbergA}. We see that these contributions can be added to terms with the same dependence in the momentum, in such a way that one obtains effective values of the LECs that are shifted $L_i\to \hat{L}_i= L_i+\Delta L_i^V$ at energies below the vector meson mass. Therefore, integrating out the massive vector boson will shift the LECs of the chiral Lagrangian \eqref{eq:pionLag} to
\begin{equation}
 \Delta L_2^V=-\Delta L_1^V=\frac{\f^4}{4}\frac{g_{V\pi\pi}^2}{m_V^4} \ .
\end{equation}

Furthermore, one could have several massive vector bosons $\bm{V}_{i\,\mu}$, $i=1,2,\ldots$ with similar couplings to the pions, in this case the shift in the LECs will have contributions from all of them
\begin{equation}
 \Delta L_2^V=-\Delta L_1^V=\frac{\f^4}{4}\sum_i \frac{g_{V_i\pi\pi}^2}{m_{V_i}^4} \ .
\end{equation}
We could be in a situation in which the LECs in the pion action vanish or are much smaller than the vector boson contribution. In this case, after integrating out the vector bosons, the LECs would automatically satisfy the relation $\hat{L}_2= -\hat{L}_1$, which corresponds to the Skyrme model \cite{Skyrme:1961vq,Skyrme:1961vr,Skyrme:1962vh}. An $O(p^4)$ coupling between the vector boson and the pions 
\begin{equation}
\sim z_4 \, \bm{V}^{\mu\nu}\cdot  \partial_\mu \bm{\pi}\times \partial_\nu \bm{\pi}\ , 
\end{equation}
does not modify this relation, see (4.38) of \cite{Harada:2003jx}, where $z_4$ is introduced in (4.27).

%%%%%%%%%%%%%%%%%%%%%%%%%%%%%%%%%%%%%%%%%%%%%
%%%%%%%%%%%%%%%%%%%%%%%%%%%%%%%%%%%%%%%%%%%%%
\section{Holographic model}\label{sec:WSS}
%%%%%%%%%%%%%%%%%%%%%%%%%%%%%%%%%%%%%%%%%%%%%
%%%%%%%%%%%%%%%%%%%%%%%%%%%%%%%%%%%%%%%%%%%%%

In this section we will introduce our string theory setup that resembles QCD and within which we can derive the pion scattering amplitude at strong coupling.
One of the closest holographic cousins to QCD is the Witten-Sakai-Sugimoto model (WSS) \cite{Witten:1998zw,Sakai:2004cn,Sakai:2005yt}; for a review see~\cite{Rebhan:2014rxa}. The background geometry is the near-horizon limit of a nonsupersymmetric $(3+1)$-dimensional D-brane intersection of $\Nc$ D$4$-branes and $\Nf$ D$8$-$\overline{\text{D}8}$-brane pairs, arranged as in table~\ref{tab:WSS}.
\begin{table}[h!]
$$
\begin{array}{c|cccccccccc}
 & 0 & 1 & 2 & 3 & \tau & 5 & 6 & 7 & 8 & 9 \\ \hline
D4 & \times & \times & \times & \times & \times & \cdot & \cdot & \cdot & \cdot & \cdot \\
D8 & \times & \times & \times & \times & \cdot &\times & \times & \times & \times & \times \\
\overline{D8} & \times & \times & \times & \times & \cdot &\times & \times & \times & \times & \times 
\end{array}
$$
\caption{D-brane intersection in the WSS model. Branes are extended along the directions marked with $\times$.}\label{tab:WSS}
\end{table}
The $\tau$ direction is compactified along a circle with SUSY-breaking boundary conditions, and the D8- and $\overline{\text{D}8}$-branes sit at separate points. Reducing along the $\tau$ direction one ends with a $(3+1)$-dimensional theory that flows at low energies to $SU(\Nc)$ QCD with $\Nf$ massless quarks when the 't Hooft coupling $\lambda_{\text{YM}}=g_{\text{YM}}^2 \Nc$ is small. When the coupling is large there is no separation of scales between the QCD sector and the five-dimensional Kaluza-Klein modes on the circle. While this latter case corresponds to the regime where the classical supergravity approximation is valid in the holographic dual, it will still be useful to use the holographic dual to study those low-energy observables that do not directly involve the Kaluza-Klein states.

In the string frame the type IIA supergravity background sourced by the D$4$-branes at zero temperature when $\Nc\to \infty$ is
\begin{equation}
\begin{split}\label{eq:D4metric}
&ds_{10}^2=\left(\frac{U}{R}\right)^{3/2}\left( \eta_{\mu\nu} dx^\mu dx^\nu +f(U)d\tau^2\right)+\left(\frac{R}{U}\right)^{3/2}\left( \frac{dU^2}{f(U)}+U^2d\Omega_4^2\right) \\
&f(U)=1-\frac{U_{\text{KK}}^3}{U^3}\ , \ e^\phi=g_s\left(\frac{U}{R}\right)^{3/4} \ , \ F_4=dC_3 =\frac{2\pi \Nc}{V_4} \epsilon_4 \ ,
\end{split}
\end{equation}
where $R^3=\pi g_s \Nc l_{\text{s}}^3$, and the Kaluza-Klein mass is $M_{\text{KK}}=(3/2) U_{\text{KK}}^{1/2}/R^{3/2}$. In our convention the $\tau$ direction has periodicity $2\pi/M_{\text{KK}}$ and $\epsilon_4$ is the volume form of a unit $S^4$, of volume $V_4=8\pi^2/3$.  The factor $f(U)$ in the metric implies that the circle collapses to zero size at $U=U_{\text{KK}}$, where the geometry ends. The $(\tau,U)$ part of the geometry can then be visualized as a `cigar', with $U=U_{\text{KK}}$ the tip. Following the usual holographic dictionary, $U$ corresponds to an energy scale of the theory in such a way that a lower bound for $U\geq U_{\text{KK}}$ sets a minimal energy scale for states in the dual field theory, thus encoding the mass gap of Yang-Mills in the confining phase. The map to field theory quantities of the parameters in the D$4$-brane geometry is
\begin{equation}\label{eq:holomap}
 R^3=\frac{1}{2} \frac{\lambda_{\text{YM}} l_{\text{s}}^2}{M_{\text{KK}}}\ ,\ U_{\text{KK}}=\frac{2}{9}\lambda_{\text{YM}} M_{\text{KK}} l_{\text{s}}^2\ ,  \ g_s \Nc=\frac{1}{2\pi} \frac{\lambda_{\text{YM}}}{M_{\text{KK}}l_{\text{s}}} 
\end{equation}
with $l_{\text{s}}$ the string length and $\lambda_{\text{YM}}=g_{\text{YM}}^2 \Nc$ the 't Hooft coupling of the ($3+1$)-dimensional dual Yang-Mills theory.

\begin{figure}[thb!]
\begin{center}
\includegraphics[scale=0.7]{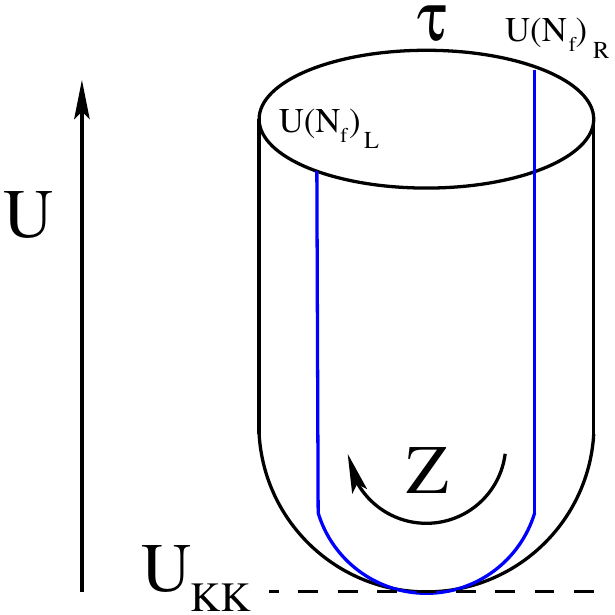}
\caption{Embedding of the D8-branes in the Witten background. We only consider massless pions in the zero temperature confining phase, which corresponds to configurations where D8- and $\overline{\text{D}8}$-branes are fixed at antipodal points of the $\tau$-direction in the asymptotic boundary and they join at a holographic radial position $U_{\text{KK}}$ where the geometry smoothly closes off.}\label{fig:SSembedding}
\end{center}
\end{figure}

Flavors are included by introducing $\Nf$ probe D$8$-branes in the geometry induced by D$4$-branes \cite{Karch:2002sh,Sakai:2004cn}. In the near-horizon limit where the D$4$-branes are replaced by the geometry displayed above, each D$8$ and $\overline{\text{D}8}$, that sit at separated points in the $\tau$ direction as $U\to\infty$, join at a finite value of the radial coordinate and form a single object. This is the geometric realization of the formation of a quark condensate and chiral symmetry breaking. The asymptotic separation between the D$8$ and $\overline{\text{D}8}$ can be changed, producing different values of the quark condensate. In this paper we will consider the simplest scenario where the D$8$- and $\overline{\text{D}8}$-branes sit at antipodal points in $\tau$ and join at the tip of the cigar. Quark masses may be introduced for non-antipodal embeddings \cite{Bergman:2007pm}; we will discuss this important generalization in section~\ref{sec:discussion}. The illustration of the D8-$\overline{\text{D}8}$ embeddings in the cigar geometry is included in Fig.~\ref{fig:SSembedding}.

The action for the D$8$-branes in the string frame
\be
 S_{\text{D}8}  = S_{\text{DBI}}+S_{\text{WZ}} \label{eq:D8action}
\ee
consists of Dirac-Born-Infeld (DBI) and Wess-Zumino (WZ) actions
\bea
 S_{\text{DBI}} & = & -T_8\int_{D8} d^9 x\,e^{-\phi}\,\textrm{STr}\sqrt{-\det\left(G_{MN}+2\pi\alpha'\,F_{MN}\right)} \\
 S_{\text{WZ}}  & = & \frac{1}{3!(2\pi)^3}\int_{D8}\, C_{3}\wedge \textrm{STr}\,F^{3} \ ,
\eea 
where $T_8=1/((2\pi)^8l_{\text{s}}^9)$ is the tension of the D$8$-brane. $F_{MN}$ is the field strength of the non-Abelian $U(\Nf)$ gauge field living on the brane, while $G_{MN}$ is the induced metric $G_{MN}=g_{\mu\nu}(X)\partial_M X^\mu \partial_N X^\nu$, with $g_{\mu\nu}$ the 10d background metric (\ref{eq:D4metric}). The embedding functions $X^\mu$ are also $\Nf\times \Nf$ Hermitian matrices, but we will take them to be proportional to the identity matrix in the following as this is sufficient for our purposes. $\textrm{STr}$ stands for symmetrized trace, this prescription for the trace is unambiguous up to $O(F^4)$ in the gauge fields \cite{Tseytlin:1997csa,Hashimoto:1997gm,Myers:1999ps,Koerber:2002zb}. At higher orders there can be additional corrections in $\alpha'=l_{\text{s}}^2$, but fortunately we will not need them. The WZ term captures the axial anomaly of the dual field theory \cite{Sakai:2004cn}. In the following we will take $\Nf=2$, so that the purely non-Abelian contributions to the WZ term vanish, corresponding to the absence of a $SU(2)^3$ axial anomaly in the dual field theory. 

The Abelian and non-Abelian components of the gauge field on the D$8$-branes are split according to
\begin{equation}
 A_M=a_M\frac{\mathbb{1}_2}{2}+A_M^a\frac{\sigma^a}{2}\ .
\end{equation}
We will denote the Abelian field strength as $f_{\mu\nu}=\partial_\mu a_\nu-\partial_\nu a_\mu$. For the non-Abelian part, we will distinguish between the linear and non-linear part of the field strength as follows
\begin{equation}
F_{\mu\nu}^a=f_{\mu\nu}^a+\epsilon^{abc}A_\mu^b A_\nu^c\ ,\ \ f_{\mu\nu}^a=\partial_\mu A_\nu^a-\partial_\nu A_\mu^a \ .
\end{equation}

For an antipodal D$8$-brane embedding it is convenient to introduce a change of coordinates $(\tau,U)\to (y,z)$ with
\begin{equation}
y=r\cos\theta \ ,  \ z=r\sin\theta \ , \ \theta=M_{\text{KK}}\tau\ , \ U^3=U_{\text{KK}}^3+U_{\text{KK}}r^2 \ .
\end{equation}
The D$8$-brane will be localized at $y=0$ and extended along the $z$ direction. The induced metric on the D$8$-brane in this case reads
\begin{equation}
 ds_8^2=\frac{4}{9}\left(\frac{R}{U_z} \right)^{3/2}\frac{U_{\text{KK}}}{U_z} dz^2+\left(\frac{U_z}{R} \right)^{3/2}\eta_{\mu\nu}dx^\mu dx^\nu +R^{3/2}U_z^{1/2}d\Omega_4^2 \ ,
\end{equation}
where
\begin{equation}
 U_z^3=U_{\text{KK}}^3+U_{\text{KK}}z^2 \ .
\end{equation}

It will be convenient to introduce dimensionless coordinates $Z=z/U_{\text{KK}}$, $x^\mu=\hat{x}^\mu/M_{\text{KK}}$ and define
\begin{equation}
 u(Z)=(1+Z^2)^{1/3} \ .
\end{equation}
Then, the induced metric in the dimensionless quantities is $G_{MN}=L^2 \hat{G}_{MN}$, where
\begin{equation}\label{eq:D8metric}
 d\hat{s}_8^2= u(Z)^{1/2}\left[\frac{dZ^2}{u(Z)^3}+ u(Z)\eta_{\mu\nu}d\hat{x}^\mu d\hat{x}^\nu +\frac{9}{4}d\Omega_4^2\right]\ , \ L^2=\frac{4}{9}R^{3/2}U_{\text{KK}}^{1/2}=\frac{4}{27}\lambda_{\text{YM}} l_{\text{s}}^2 \ .
\end{equation}
The DBI action can be split according to the factors of the field strength, expanding up to $O(F^4)$,
\begin{equation}\label{eq:SDBI}
S_{\text{DBI}}=-\tT_8 \int d^4 \hat{x} dZ\,  u(Z)^2\left(2+\left(\frac{\pi\alpha'}{L^2}\right)^2\cL_{\text{DBI}}^{[2]}+\left(\frac{\pi\alpha'}{L^2}\right)^4 \cL_{\text{DBI}}^{[4]}+\ldots\right) \ ,
\end{equation}
where
\begin{equation}\label{eq:tensionD8}
 \tT_8=\frac{3}{2g_s} V_4 R^3 L^6 T_8=\frac{\Nc\lambda_{\text{YM}}^3}{3^9 \pi^5} \ , \ \frac{\pi \alpha'}{L^2}=\frac{27 \pi}{4\lambda_{\text{YM}}} \ .
\end{equation}
The Lagrangian densities, in terms of dimensionless gauge fields, coordinates, and the metric read
\begin{equation}\label{eq:DBIaction}
\begin{split}
&\cL_{\text{DBI}}^{[2]}=\frac{1}{2} F_{MN}^a F^{a\,MN}+\frac{1}{2} f_{MN} f^{MN} \\
&\cL_{\text{DBI}}^{[4]}=-\frac{1}{6}\left[F^{a\, MA} F^a_{NA} F^b_{MB} F^{b\, NB}+\frac{1}{2} F^{a\, MN} F^{a\, AB} F^b_{MB}F^b_{AN}\right. \\
& \qquad\qquad\qquad\left. -\frac{1}{8} \left(F^{a\, MN}F^a_{MN} F^{b\, AB} F^b_{AB}+2F^{a\, MN} F^{a\, AB} F^b_{MN}F^b_{AB} \right)\right]+O(f^2F^2,f^4)\ .
\end{split}
\end{equation}
In the quartic action (\ref{eq:DBIaction}) we omitted writing the explicit form of the Abelian and mixed terms since at tree level they do not contribute to a quartic interaction with non-Abelian fields in the external legs.

The Wess-Zumino action is non-vanishing due to the background four-form
\begin{equation}
 \frac{1}{2\pi}\int_{S^4} F_4=\Nc 
\end{equation}
in such a way that the Wess-Zumino action is proportional to a five-dimensional Chern-Simons term for the gauge fields on the brane
\begin{equation}\label{eq:WZaction}
 S_{\text{WZ}}=\frac{\Nc}{24\pi^2}\int_{M^4\times \mathbb{R}} \omega_5(A)=\int d^4\hat{x} dZ\, \cL_{\text{WZ}} \ . 
\end{equation}
For purely non-Abelian $SU(2)$ fields, the Chern-Simons term vanishes, as mentioned before. However, there are mixed terms between Abelian and non-Abelian components
\begin{equation}\label{eq:WZlag}
 \cL_{\text{WZ}}=\frac{\Nc}{32\pi^2} \epsilon^{MNLPQ} a_M \partial_N A_L^a\partial_P A_Q^a+ O(A^4 a) \ .
\end{equation}

%%%%%%%%%%%%%%%%%%%%%%%%%%%%%%%%%%%%%%%%%%%%%
\subsection{Chiral symmetry and pion mode}
%%%%%%%%%%%%%%%%%%%%%%%%%%%%%%%%%%%%%%%%%%%%%

Let us discuss the realization of chiral symmetry and the existence of a massless mode corresponding to the pion. Considering only fields that are constant on the $S^4$, there are two boundary values that have to be specified for the gauge field on the D$8$-branes, each of them mapping to background values for the $U(\Nf)_L$ and $U(\Nf)_R$ gauge fields 
\begin{equation}\label{eq:holosources}
 \lim_{Z\to +\infty} A_\mu(x,Z)=L_\mu(x)=V_{\mu}(x)+A_{5\,\mu}(x)\ , \ \lim_{Z\to -\infty} A_\mu(x,Z)=R_\mu(x)=V_\mu(x)-A_{5\,\mu}(x) \ ,
\end{equation}
where $V_\mu$ are the $U(\Nf)_V$ (vector) and $A_{5\,\mu}$ the $U(\Nf)_A$ (axial) gauge fields. The set of gauge transformations $U(x,Z)$ of the field on the D$8$-branes
\begin{equation}
 A_M =U^{-1}A_M U-i U^{-1}\partial_M U \ ,
\end{equation}
generate gauge transformations of the background left- and right-handed gauge fields in the dual field theory
\begin{equation}
\lim_{Z\to +\infty} U(x,Z)=U_L(x) \ , \ \lim_{Z\to -\infty} U(x,Z)=U_R(x) \ .
\end{equation}
In the $A_Z=0$ gauge the set of allowed bulk gauge transformations is reduced to $Z$-independent transformations $U_{L,R}(x)$, and the global subgroup are constant transformations. These global transformations are identified with the $U(\Nf)_L\times U(\Nf)_R$ flavor group of the dual field theory. Note that, when the sources for the flavor currents are turned off $\lim_{Z\to\pm \infty} A_\mu=0$, the global transformations do not change the boundary values of the gauge field as expected. As we mentioned, chiral symmetry is spontaneously broken $U(\Nf)_L\times U(\Nf)_R\to U(\Nf)_V$ in the dual field theory, so there should be a mode on the D$8$-brane that corresponds to massless pions. We will elucidate this below.

From the quadratic action in \eqref{eq:DBIaction} we obtain the following set of linearized equations,
\begin{equation}\label{eq:lineareqs}
\begin{split}
&\partial_Z\left[u(Z)^3 f^a_{Z\mu} \right]+\frac{1}{u(Z)}\eta^{\alpha\beta}\partial_\alpha f^{a}_{\beta \mu}= 0 \\
&\eta^{\alpha\beta}\partial_\alpha f^{a}_{\beta Z}=0 \ ,
\end{split}
\end{equation}
where $f_{MN}^a=\partial_M A_N^a-\partial_N A_M^a$.  We will split the gauge potential in transverse, longitudinal, radial, and gauge parts:
\begin{equation}
A_\mu^a=A_\mu^{\perp\, a}+\partial_\mu A^{\parallel\, a}+\partial_\mu C^a\ , \ A_Z^a=B_Z^a+\partial_Z C^a\ , \ \eta^{\mu\nu}\partial_\mu A_\nu^{\perp\, a}=0 \ .
\end{equation}
The second equation in \eqref{eq:lineareqs} imposes the conditions ($\partial^2=\eta^{\alpha\beta}\partial_\alpha\partial_\beta$)
\begin{equation}\label{eq:BZsol}
\partial^2 B_Z^a=\partial^2 \partial_Z A^{\parallel\, a}\ .
\end{equation}
Then, either $\partial^2 A_M^a=0$, or $B_Z$ and $A^{\parallel\, a}$ are pure gauge and can be absorbed in $C^a$. In the case when the first condition is true, corresponding to a massless mode, the first equation in \eqref{eq:lineareqs} becomes
\begin{equation}
\partial_Z\left[u(Z)^3 (\partial_Z A_\mu^{a}-\partial_\mu A_Z^a) \right]= 0 \ .
\end{equation}
The solutions are, up to gauge transformations $C^a(x,Z)$,
\begin{equation}\label{eq:linearsolWSS}
A_\mu^{a}(x,Z)=\hat{V}_\mu^{a}(x)+\hat{A}_{5\,\mu}^{a}(x)\frac{2}{\pi}\arctan(Z) \ , \ A_Z^a(x,Z)=2\varphi^a(x)\phi_0(Z)=\frac{2}{\pi}\frac{\varphi^a(x)}{1+Z^2} \ .
\end{equation}
If we set $\hat{V}_\mu^a=\hat{A}_{5\,\mu}=0$ the mode is normalizable, so there is a massless particle in the dual field theory, which will be identified with the pion. This in fact is an exact solution of the $O(F^2)$ action. The normalizable solution has the field strength
\begin{equation}\label{eq:solpion}
F_{\mu Z}^a=2 \partial_\mu \varphi^a \phi_0(Z) \ , \ F_{\mu\nu}^a=0 \ .
\end{equation}
Let us show now that indeed $\varphi^a$ is the pion field up to an overall normalization. Following the usual prescription of the holographic dictionary, we first compute the canonical momentum conjugate to the gauge field
\begin{equation}
 \pi_a^\mu=\frac{\delta S_{D8}}{\delta\left(\partial_Z A_\mu^a\right)} \ .
\end{equation}
From \eqref{eq:SDBI},
\begin{equation}\label{eq:canmom}
 \pi_a^{\mu}= -\tilde{T}_8\left(\frac{\pi\alpha'}{L^2}\right)^{2} u(Z)^2\left[ \Pi_a^{[2]\,\mu} +\left(\frac{\pi\alpha'}{L^2}\right)^{2} \Pi_a^{[4]\, \mu} +\ldots\right] \ ,
\end{equation}
where, using \eqref{eq:DBIaction},
\begin{equation}
\begin{split}
& \Pi_a^{[2]\,\mu}=2F^{a\,Z\mu} \\
& \Pi_a^{[4]\,\mu}=\frac{1}{6}\left[4F_{L\nu }^b\left(F^{a\,\mu L}F^{b\,Z \nu}+F^{b\,\mu L}F^{a\,Z \nu}\right)+4F_{\,L\nu}^a F^{b\,\mu L}F^{b\,Z \nu}\right. \\
&\qquad\qquad\qquad\left.+\left(F^{a\,Z\mu}F_{NL}^b F^{b\,NL}+2F^{b\,Z\mu}F^{b\,NL}F_{\,NL}^a\right)\right] \ .
\end{split}
\end{equation}
The expectation values of the flavor currents $J_L^{a\,\mu}$ and $J_R^{a\,\mu}$ are obtained from the boundary values of the canonical momentum\footnote{The relative sign stems from the variation of the D$8$-brane on-shell action $\delta S_{on-shell}=\int d^4 x \left(\vev{J_L^{\mu\,a}}\delta L_\mu^a+\vev{J_R^{\mu\,a}}\delta R_\mu^a\right)$, with the left current at the upper limit of the radial integration and the right current at the lower limit.}
\begin{equation}\label{eq:vevspi}
 \vev{J_L^{\mu\,a}}=\lim_{Z\to +\infty} \pi_a^\mu \ , \ \vev{J_R^{\mu\,a}}=-\lim_{Z\to -\infty} \pi_a^\mu \ .
\end{equation}
To leading order in the field, and restoring units in the $x^\mu$ directions,
\begin{equation}
 \vev{J_5^{\mu\,a}}=\vev{J_L^{\mu\,a}}-\vev{J_R^{\mu\,a}}\simeq \f^2\eta^{\mu\nu}\partial_\nu \varphi^a\ , \ \vev{J_V^{\mu\,a}}=\vev{J_L^{\mu\,a}}+\vev{J_R^{\mu\,a}}\simeq 0 \  ,
\end{equation}
where $\f^2$ is given in \eqref{eq:coefsSS}. This shows that $\varphi^a$ is proportional to the pion field, since the axial current is proportional to its gradient. Our next goal is to find the pion effective action. Usually this has been done by identifying $\varphi^a$ as the pion field and using the D$8$-brane action integrated over the holographic radial coordinate $Z$ as the effective action for the field $\varphi^a$. Within this general idea there are two different approaches, an {\em off-shell} approach where $\varphi^a$ is taken to be an arbitrary function and an {\em on-shell} approach where it is a massless field $\partial^2\varphi^a=0$. The off-shell approach is the one used originally in the WSS model \cite{Sakai:2004cn,Sakai:2005yt}, and has been also employed in other phenomenological models  \cite{Hirn:2005nr,DaRold:2005vr,Erlich:2005qh,Chivukula:2006fxj,Panico:2007qd,Domenech:2010aq,Colangelo:2012ipa,Harada:2014lza,Espriu:2020ise}, while the on-shell approach was introduced in \cite{Domokos:2014ura} for the AdS/QCD model of \cite{Hirn:2005nr}. 

%%%%%%%%%%%%%%%%%%%%%%%%%%%%%%%%%%%%%%%%%%%%%
\subsection{Effective action for the pion field and vector mesons}
%%%%%%%%%%%%%%%%%%%%%%%%%%%%%%%%%%%%%%%%%

The basic idea of the off-shell approach is to expand the field in ``Kaluza-Klein'' modes of the holographic radial direction, which, excluding the massless mode, are of the form
\begin{equation}
 A_\mu(x,Z)=\sum_n A_\mu^{(n)}(x) \psi_n(Z) \ ,  \ A_Z^{(n)}=0 \ .
\end{equation}
Here the functions $\psi_n$ are eigenfunctions of the radial derivative part of the equations
\begin{equation}
u(Z)\partial_Z\left[u(Z)^3 \partial_Z \psi_n(Z)\right]= -m_n^2 \psi_n(Z) \ ,  \ \lim_{Z\to \pm \infty} \psi_n(Z)=0 \ ,
\end{equation}
where $m_n^2$ determine the masses of mesons in the dual field theory. Then, from \eqref{eq:lineareqs}
\begin{equation}\label{eq:KKeqs}
 \partial^2 A_\mu^{(n)}-m_n^2 A_\mu^{(n)}= 0 \ , \ \eta^{\alpha\beta}\partial_\alpha A_\beta^{(n)}=0 \ .
\end{equation}
Introducing the Kaluza-Klein expansion, together with the massless solution back in the action \eqref{eq:SDBI} and integrating over the radial coordinate results in an action for the 4D fields $A_\mu^{(n)}$ and $\varphi$. This is to be interpreted as the effective action for the meson fields in the dual field theory, with interactions determined by non-quadratic terms. In this derivation the 4D fields are off-shell, {\emph{i.e.}}, the equations of motion \eqref{eq:KKeqs} are not imposed. In previous derivations only the $O(F^2)$ terms were kept, while higher $\alpha'$ corrections to the DBI action were neglected. This restricts the terms in the action to be at most quartic in the fields (for pions and vector mesons) and to have at most four derivatives in the field theory directions. Terms $O(F^4)$ can contribute at the same order of derivatives and fields, so they must be included if one is interested in computing the value of the LECs at finite 't Hooft coupling.

The off-shell Kaluza-Klein expansion is essentially the approach applied to the WSS construction in \cite{Sakai:2004cn,Sakai:2005yt} to derive the meson effective action. The action \eqref{eq:actionWSS} was actually computed in a slightly different way. Allowing the boundary conditions of the gauge field to be fixed only up to boundary gauge transformations, it is possible to apply a gauge transformation to the solution \eqref{eq:linearsolWSS} (with $\hat{V}_\mu=\hat{A}_{5\,\mu}=0$) such that the radial component vanishes and the pion field $\varphi^a$ is moved to the components of the gauge potential along the field theory directions. Beyond the linear approximation, this amounts to fixing $A_Z=0$ and introducing instead a pure gauge configuration for the boundary gauge fields, that is taken to be
\begin{equation}\label{eq:pionsolPots}
\hat{V}_\mu=-\frac{i}{2}\Sigma^{-1}\partial_\mu \Sigma \ , \ \hat{A}_{5\,\mu}=-\frac{i}{2}\Sigma^{-1}\partial_\mu \Sigma \ .
\end{equation}
We then identify $\Sigma(x)$ as the $SU(\Nf)$ matrix of the pions. In this case the field strengths are
\begin{equation}\label{eq:solpionWSS}
F_{Z\mu}=-i\Sigma^{-1}\partial_\mu \Sigma\, \phi_0(Z) \ , \ F_{\mu\nu}=-i\left[\Sigma^{-1}\partial_\mu \Sigma,\Sigma^{-1}\partial_\nu \Sigma \right](\psi_0(Z)-1)\psi_0(Z)
\end{equation}
with 
\begin{equation}\label{eq:psi0}
 \psi_0(Z)=\frac{1}{2}+\frac{1}{\pi}\arctan(Z) \ , 
\end{equation}
and we have used $\psi_0'(Z)=\phi_0(Z)$. Plugging \eqref{eq:solpionWSS} back in the action \eqref{eq:SDBI} and keeping only $O(F^2)$ terms one finds \eqref{eq:actionWSS} up to quartic order in the fields. The terms that involve only the pion have the form \eqref{eq:pionaction} with
\begin{equation}\label{eq:coefsSS}
\begin{split}
 &\f^2=8M_{\text{KK}}^2 \frac{\Nc\lambda_{\text{YM}}}{4^2 3^3 \pi^3}  \int_{-\infty}^\infty dZ u^3\phi_0^2=\frac{\Nc\lambda_{\text{YM}}}{54 \pi^4}M_{\text{KK}}^2 \\ 
 & L_2^{SU(2)}= -L_1^{SU(2)}=2\frac{\Nc\lambda_{\text{YM}}}{4^2 3^3 \pi^3}\int_{-\infty}^\infty dZ \frac{(\psi_0-1)^2\psi_0^2}{u}=\frac{\Nc\lambda_{\text{YM}}}{ 6^3 \pi^7}b \ ,\ b\approx 15.25 \ .
\end{split}
\end{equation}
One should note that since the boundary values of the gauge potentials do not vanish, this actually corresponds to having a non-zero source proportional to the derivatives of the pion field, in such a way that the pion field enters as a ``spurion''. In order to demonstrate this we will introduce external gauge fields $\hat{\cal V}_\mu$, $\hat{\cal A}_{5\,\mu}$ coupled to the axial and vector currents by modifying the boundary conditions in \eqref{eq:pionsolPots}
\begin{equation}\label{eq:pionsolPots2}
\hat{V}_\mu=\Sigma^{-1}\hat{\cal V}_\mu\Sigma-\frac{i}{2}\Sigma^{-1}\partial_\mu \Sigma \ , \ \hat{A}_{5\,\mu}=\Sigma^{-1}\hat{\cal A}_{5\,\mu}\Sigma-\frac{i}{2}\Sigma^{-1}\partial_\mu \Sigma \ .
\end{equation}
Then, a simultaneous transformation of the pion field and the boundary gauge fields leaves the boundary values of the bulk gauge field invariant. 

Expanding up to quartic order in the fields, the shift \eqref{eq:mesonshift} removes the source terms depending on the pion field from the effective action and as we have discussed in the introduction, terms quartic in the pion field go away, so the only contributions depending on the pion field left are quadratic or interaction terms involving vector mesons. That the spurion action is able to capture the LECs can be understood from the fact that external gauge fields should be dressed by the physical pion field (once massive vector mesons have been integrated out) in the same way as they are for the spurion in \eqref{eq:pionsolPots2}. However, even when the massive vector mesons have not been integrated out, there are terms at $O(F^4)$ in the action \eqref{eq:SDBI} not included in the original derivation \cite{Sakai:2004cn} that contribute at the same order in fields (quartic) and derivatives (four), albeit they are relatively suppressed by a $1/\lambda_{\text{YM}}^2$ factor. 

Moving on to the on-shell approach, it has not really been applied to the WSS model, but to other AdS/QCD models with an IR cutoff in the holographic radial direction \cite{Domokos:2014ura}. In this case the pion field is typically identified with the value of the gauge field at the cutoff, which together with a fixed asymptotic value at the boundary determines the solution for the gauge field. Then one proceeds in a similar way as in the off-shell derivation, evaluating the action on the solution and integrating over the radial direction to obtain the effective action. However, there are two main differences with the off-shell approach. The first one is that there is no expansion in Kaluza-Klein modes. Instead, solutions are found by fixing the value of the field at the cutoff, so even for the linearized equations they will typically consists of a superposition of the massless mode and the whole Kaluza-Klein tower. The second difference is that the full set of equations is solved, including equations with only field theory derivatives and non-linear terms. This can be done systematically using an expansion in derivatives and factors of the pion field, which is on-shell in this derivation ({\emph{i.e.}}, terms proportional to the equations of motion of the pion field vanish). 

It was also argued in \cite{Domokos:2014ura}, and shown for the AdS/QCD model introduced in \cite{Hirn:2005nr}, that the off-shell and on-shell derivations of the effective action should agree if the former is put on-shell which in the low momentum expansion requires integrating out all the massive vector bosons. In the WSS model one could in principle attempt a similar on-shell derivation contemplating the pion as the value of the field at an IR cutoff at $Z=0$.

Our approach using scattering amplitudes have some similarities with the on-shell approach in that we will be solving the equations of motion in an expansion around the linearized solution, however, it will be the UV rather than the IR value of the gauge fields that will determine the expansion.

%%%%%%%%%%%%%%%%%%%%%%%%%%%%%%%%%%%%%%%%%%%%%
%%%%%%%%%%%%%%%%%%%%%%%%%%%%%%%%%%%%%%%%%%%%%
\section{Holographic calculation of the pion scattering amplitude}\label{sec:scattering}
%%%%%%%%%%%%%%%%%%%%%%%%%%%%%%%%%%%%%%%%%%%%%
%%%%%%%%%%%%%%%%%%%%%%%%%%%%%%%%%%%%%%%%%%%%%

The holographic dictionary instructs to map gauge-invariant operators to fields in the gravity dual. The pion should be understood as a mode produced by the axial current operator. We can thus obtain the pion scattering amplitudes from axial current correlators via an LSZ reduction formula where we have to identify the massless poles appearing in the correlators. The pion propagator will be determined by the two-point function of the axial current and the $2\to 2$ scattering amplitude henceforth by the four-point function.

From \eqref{eq:vevspi}, the expectation value (vev) of the axial current can be computed from the asymptotic values of the canonical momentum conjugate to the gauge fields 
\begin{equation}\label{eq:vevJ5}
\vev{J_{5\,a}^\mu}=\left(\lim_{Z\to \infty} + \lim_{Z\to -\infty}\right)\pi_a^\mu \ .
\end{equation}
We can extract higher order correlators from the vev by taking variations with respect to an external axial gauge field
\begin{equation}
\vev{J_5^{\mu \,a}(x) J_5^{\mu_1\,a_1}(y_1)\cdots J_5^{\mu_n\, a_n}(y_n)}=(-i)^n\prod_{i=1}^n \frac{\delta}{\delta A_{5\,\mu_i}^{a_i}(y_i)}\vev{J_{5\,a}^\mu(x)}\ ,
\end{equation}
where the external gauge fields are identified as the asymptotic values of the gauge fields \eqref{eq:holosources}. Taking into acount that 
\begin{equation}
\int d^4 x A_{5\,\mu}^a(x) J_5^{\mu\, a}(x)=\int \frac{d^4 q}{(2\pi)^4}  A_{5\,\mu}^a(-q) J_5^{\mu\, a}(q) \ ,
\end{equation}
the analogous formula in momentum space reads
\begin{equation}
\vev{J_5^{\mu \,a}(p) J_5^{\mu_1\,a_1}(q_1)\cdots J_5^{\mu_n\, a_n}(q_n)}=(-i)^n\prod_{i=1}^n \frac{\delta}{\delta A_{5\,\mu_i}^{a_i}(-q_i)}\vev{J_{5\,a}^\mu(p)} \ .
\end{equation}

%%%%%%%%%%%%%%%%%%%%%%%%%%%%%%%%%%%%%%%%%%%%%
\subsection{Expansion in a background axial gauge field}
%%%%%%%%%%%%%%%%%%%%%%%%%%%%%%%%%%%%%%%%%%%%%

We will set $V_\mu^a=0$ and $A_{5\,\mu}^a\sim O(\epsilon)$ with $\epsilon$ treated as a small parameter. The solutions for the gauge field on the D$8$-branes will be expanded in $\epsilon$, that counts the number of factors of the source appearing in each term of the expansion
\begin{equation}
 A_M=\epsilon A_M^{(1)}+\epsilon^2 A_M^{(2)}+\epsilon^3 A_M^{(3)}+\ldots \ .
\end{equation}
The $O(\epsilon)$ term contribution captures the two-point function of the current and the $O(\epsilon^3)$ contribution the four-point function. Our goal is to compute both in the following.

At each order we can arrange the equations of motion for the gauge field as follows
\begin{equation}
\begin{split}
&\partial_Z\left[u(Z)^3 f^{(n)\, a}_{Z\mu} \right]+\frac{1}{u(Z)}\eta^{\alpha\beta}\partial_\alpha f^ {(n)\,a}_{\beta \mu}= I^{(n)\, a}_\mu \\
&u(Z)^3\eta^{\alpha\beta}\partial_\alpha f^{(n)\,a}_{\beta Z}=I^{(n)\, a}_Z\ ,
\end{split}
\end{equation}
where $I^{(1)\,a}_\mu=I_Z^{(1)\, a}=0$. Following the holographic dictionary, we should impose boundary conditions such that
\begin{equation}
\lim_{Z\to \pm \infty} A_\mu^{(1)\,a}(x,Z)=\pm \hat{A}_{5\,\mu}^a(x)\ , \ \lim_{Z\to \pm \infty} A_\mu^{(n)\,a}(x,Z)=0 \ , n>1 \ .
\end{equation}
We will work with Fourier transforms of the fields, and split the gauge potentials as before in transverse, longitudinal, radial, and gauge parts 
\begin{equation}
A_\mu^{(n)\,a}(q)=A_\mu^{(n)\mu\perp\, a}+i q_\mu A^{(n)\,\parallel\, a}+i q_\mu C^{(n)\,a}\ , \ A_Z^{(n)\,a}(q)=B_Z^{(n)\,a}+\partial_Z C^{(n)\,a}\ .
\end{equation}
The equations of motion for each component of the gauge field are
\begin{equation}\label{eq:eomseps}
\begin{split}
&\partial_Z\left[u(Z)^3 \partial_Z A_\mu^{(n)\perp \, a} \right]-\frac{q^2}{u(Z)} A^ {(n)\,\perp\,a}_\mu= I^{(n)\,\perp\, a}_\mu \\
& q^2\partial_Z\left[u(Z)^3\left( B_Z^{(n)\,a}- \partial_Z A^{(n)\parallel \, a}\right) \right]=i q^\alpha I^{(n)\, a}_\alpha \\
&q^2 u(Z)^3\left(\partial_Z A^{(n)\,\parallel\,a}-B_Z^{(n)\,a}\right)=I^{(n)\, a}_Z \ .
\end{split}
\end{equation}
Here and in the following indices will be raised and lowered with the flat Minkowski metric. At orders $n>1$ we have to solve inhomogeneous equations. Since $i q^\alpha I^{(n)\,a}_\alpha+\partial_Z I^{(n)\,a}_Z=0$ we can split
\begin{equation} 
I^{(n)\,a}_\mu=i q_\mu \partial_Z J^{(n)\,a}+I_\mu^{(n)\,\perp\,a} \ , \ I_Z^{(n)\,a}=q^2 J^{(n)\,a} \ .
\end{equation}
Then, the inhomogeneous solution for the longitudinal and radial parts is, up to gauge transformations,
\begin{equation}\label{eq:AparSol}
A^{(n)\,\parallel\,a}=0,\ \ B_Z^{(n)\,a}=-\frac{1}{u(Z)^3}J^{(n)\,a} \ .
\end{equation}
The field strengths are
\begin{equation} \label{eq:fieldst}
\begin{split}
&f^{(n)\, a}_{Z\mu}=\partial_Z A^{(n)\, \perp\,a}_\mu+\frac{i q_\mu}{q^2}\frac{I_Z^{(n)\,a}}{u^3} \\
&f^{(n)\, a}_{\mu\nu}=i\left(q_\mu A^{(n)\, \perp\,a}_\nu-q_\nu A^{(n)\, \perp\,a}_\mu\right) \ .
\end{split}
\end{equation}
For the transverse component of the gauge field, the inhomogeneous solution can be formally found using a Green's function
\begin{equation}\label{eq:AperSol}
A^{(n)\,\perp\,a}(Z)=\int_{-\infty}^\infty dZ_1\,G(Z,Z_1;q^2)I^{(n)\,\perp\, a}_\mu(Z_1) \ .
\end{equation}
The Green's function is the solution to
\begin{equation}\label{eq:Greenfunc}
\partial_Z\left[u(Z)^3 \partial_Z G(Z,Z_1;q^2) \right]-\frac{q^2}{u(Z)}G(Z,Z_1;q^2)=\delta(Z-Z_1) 
\end{equation}
with the boundary conditions
\begin{equation}
\lim_{Z\to \pm\infty} G(Z,Z_1;q^2)=0 \ .
\end{equation}
For general values of $q^2$ we have not been able to found a closed form analytic solution for the Green's function. However, for $q^2=0$ it takes a simple form
\begin{equation}
G(Z,Z_1;0)=\pi\left\{ \begin{array}{ccc} \psi_0(Z) (\psi_0(Z_1)-1) &,& Z<Z_1\\ \psi_0(Z_1)(\psi_0(Z)-1) &,& Z>Z_1 \end{array}\right. \ , 
\end{equation}
where $\psi_0(Z)$ was given in \eqref{eq:psi0}.\footnote{Although not needed in our paper, one can obtain the finite momentum result using an analytic expansion of the Green's function around $q^2=0$:
\begin{equation}\label{eq:Gqexp}
 G(Z,Z_1;q^2)=G^{(0)}(Z,Z_1)+q^2 G^{(1)}(Z,Z_1)+(q^2)^2 G^{(2)}(Z,Z_1)+\ldots \ ,
\end{equation}
where $G^{(0)}(Z,Z_1)=G(Z,Z_1;0)$ and where higher order terms in the expansion can be computed iteratively using
\begin{equation}\label{eq:Gqexpcoef}
  G^{(n)}(Z,Z_1)=\int_{-\infty}^\infty dZ_2 \, G(Z,Z_2;0) \frac{G^{(n-1)}(Z_2,Z_1)}{u(Z_2)} \ .
\end{equation}
}
 The asymptotic expansion is
\begin{equation}\label{eq:Gasymp}
G(Z,Z_1;0)\underset{|Z|\to  \infty}{\simeq}  -\frac{1}{Z}\left\{ \begin{array}{ccc}  \psi_0(Z_1)-1 &,& Z\to-\infty\\ \psi_0(Z_1)  &,& Z\to +\infty\end{array}\right. \ .
\end{equation}

%%%%%%%%%%%%%%%%%%%%%%%%%%%%%%%%%%%%%%%%%%%%%
\subsection{Two-point function of the axial current}
%%%%%%%%%%%%%%%%%%%%%%%%%%%%%%%%%%%%%%%%%%%%%

Before computing the scattering amplitude we need to compute the residue of the massless pole in the axial current two-point function. This determines the pion decay constant, that enters as well in the coefficients of the four-pion interaction, so it is necessary to know its value in order to compare with ChPT.

In order to compute the two-point function it is enough to find the solution for the D$8$-brane gauge field at $O(\epsilon)$, so the solutions to the linearized equations of motion suffice. Since we are interested in the massless pole, we can expand around $q^2=0$. Then, the solution is  \eqref{eq:linearsolWSS} plus small corrections
\begin{equation}\label{eq:order1sol}
  A_\mu^{(1)\,a}=\frac{2}{\pi}\arctan(Z)\hat{A}_{5\,\mu}^a(q)+O(q^2) \ , \ A_Z^{(1)\,a}=-2\frac{i q^\alpha}{q^2} \hat{A}_{5\,\alpha}^a(q)\phi_0(Z)+O(q^2) \ ,
\end{equation}
Here we have taken into account \eqref{eq:BZsol} for $q^2\neq 0$. The field strength at $O(\epsilon)$ is
\begin{equation}\label{eq:fZ1}
  f_{Z\mu}^{(1)\,a}=2\phi_0(Z) \left(\delta_\mu^{\nu}-\frac{ q_\mu q^\nu}{q^2}\right)\hat{A}_{5\,\nu}^a(q) \ . 
\end{equation}

The expectation value of the axial current is determined by the canonical momentum as in \eqref{eq:vevJ5}, and to $O(\epsilon)$ we only need the term originating from the $O(F^2)$ terms in the D$8$-brane action, $\Pi^{[2]}$ in \eqref{eq:canmom}. We find for the canonical momentum, using \eqref{eq:tensionD8},
\begin{equation}
 \pi_a^\mu\simeq -4\tilde{T}_8\left(\frac{\pi\alpha'}{L^2}\right)^{2}\left(\eta^{\mu\nu}-\frac{ q^\mu q^\nu}{q^2}\right)\hat{A}_{5\,\nu}^a(q) u(Z)^3 \phi_0(Z)=-\frac{\Nc\lambda_{\text{YM}}}{108 \pi^4}\left(\eta^{\mu\nu}-\frac{q^\mu q^\nu}{q^2}\right)\hat{A}_{5\,\nu}^a(q) \ .
\end{equation}
Restoring units, the expectation value of the axial current at this order reads
\begin{equation}
 \vev{J_5^{a\,\mu}(q)}\simeq-\frac{\Nc\lambda_{\text{YM}}}{54 \pi^4}M_{\text{KK}}^2\left(\eta^{\mu\nu}-\frac{q^\mu q^\nu}{q^2}\right)A_{5\,\nu}^a(q)=-\f^2\left(\eta^{\mu\nu}-\frac{q^\mu q^\nu}{q^2}\right)A_{5\,\nu}^a(q) \ .
\end{equation}
This determines the susceptibilities of the axial current. Indeed, considering configurations constant in time,
\begin{equation}
 \delta A_{5\, 0}^a(q)=\delta \mu_5^a (2\pi)\delta(q_0)\ , \ \delta \vev{J_5^{a\,0}(q)}=\delta\rho_5^a (2\pi)\delta(q_0) \ ,
\end{equation}
with $\mu_5^a$ the axial chemical potentials and $\rho_5^a$ the axial charge densities, we obtain
\begin{equation}
 \delta \rho_5^a\simeq \f^2 \delta^{ab}\delta \mu_5^b\ \ \Rightarrow\ \ \frac{\partial \rho_5^a}{\partial \mu_5^b}\simeq \f^2 \delta^{ab} \ .
\end{equation}
The two-point function of the axial current is, at this order
\begin{equation}
 \vev{J_5^{a\,\mu}(-q)J_5^{b\,\nu}(q)}\simeq i\f^2\left(\eta^{\mu\nu}-\frac{q^\mu q^\nu}{q^2}\right)\delta^{ab} \ .
\end{equation}
The residue of the massless pole agrees with the expectation from the effective action. Note that the Ward identity for current conservation is satisfied
\begin{equation}\label{eq:wardfixed}
 -iq_\mu \vev{J_5^{a\,\mu}(-q)J_5^{b\,\nu}(q)}= 0 \ .
\end{equation}
This mends the problem of the chiral effective theory Ward identity for the axial current that we mentioned before, one should had included the contact term proportional to the susceptibilities in \eqref{eq:wardJ5}.

%%%%%%%%%%%%%%%%%%%%%%%%%%%%%%%%%%%%%%%%%%%%%
\subsection{Four-point function of the axial current}\label{sec:fourpoint}
%%%%%%%%%%%%%%%%%%%%%%%%%%%%%%%%%%%%%%%%%%%%%

Before computing the scattering amplitude we need to find the leading pole contributions to the four-point function of the current, from the $O(\epsilon^3)$ terms in the expansion of the D$8$-brane gauge field. This boils down to the calculation of the $O(\epsilon^2)$ and $O(\epsilon^3)$ inhomogeneous terms in \eqref{eq:eomseps}, which are then introduced in \eqref{eq:AparSol} and \eqref{eq:AperSol} to get the solution for the gauge field. 

\begin{figure}[thb!]
\begin{center}
\includegraphics[scale=0.3]{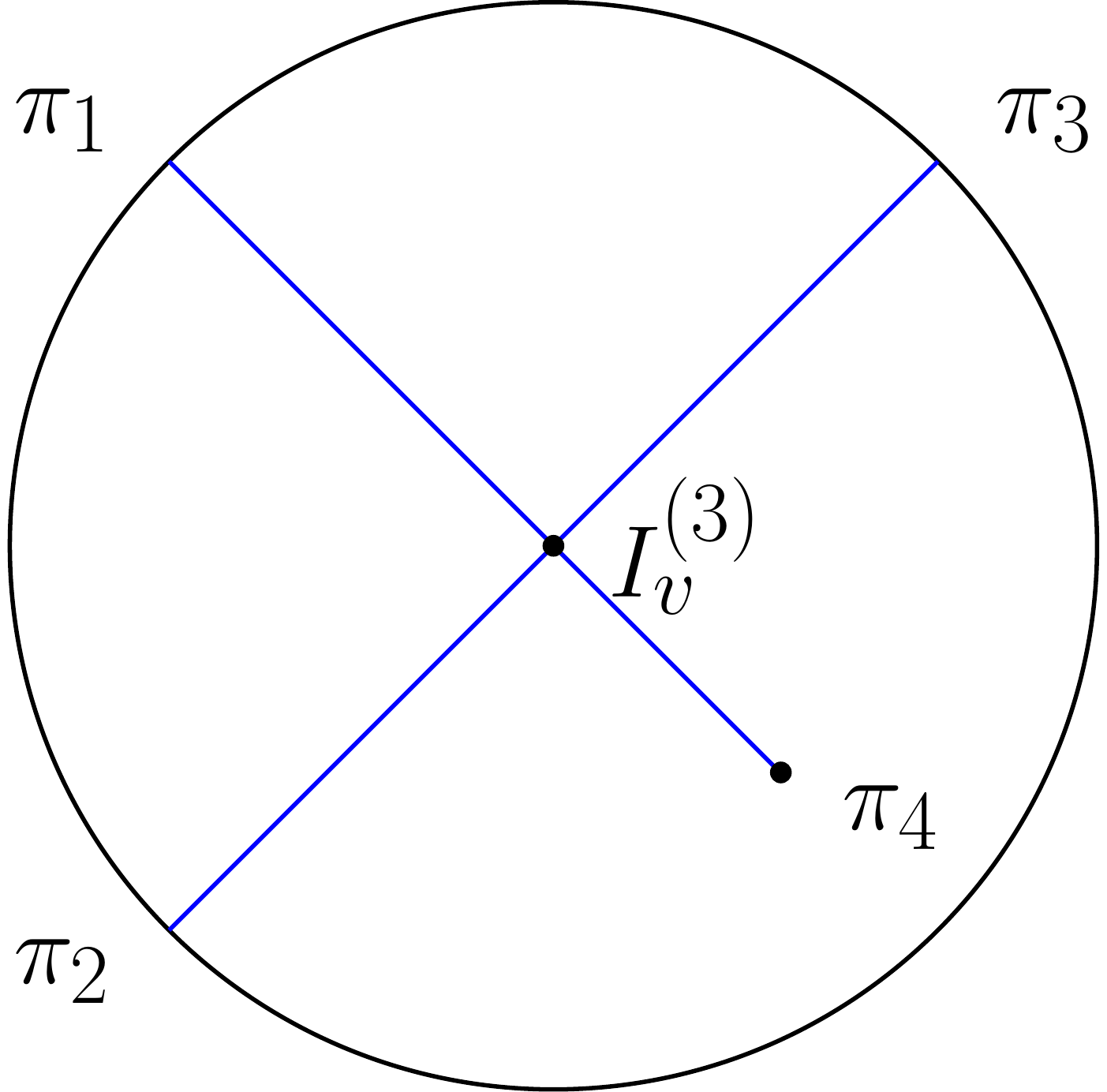}
\qquad\qquad
\includegraphics[scale=0.3]{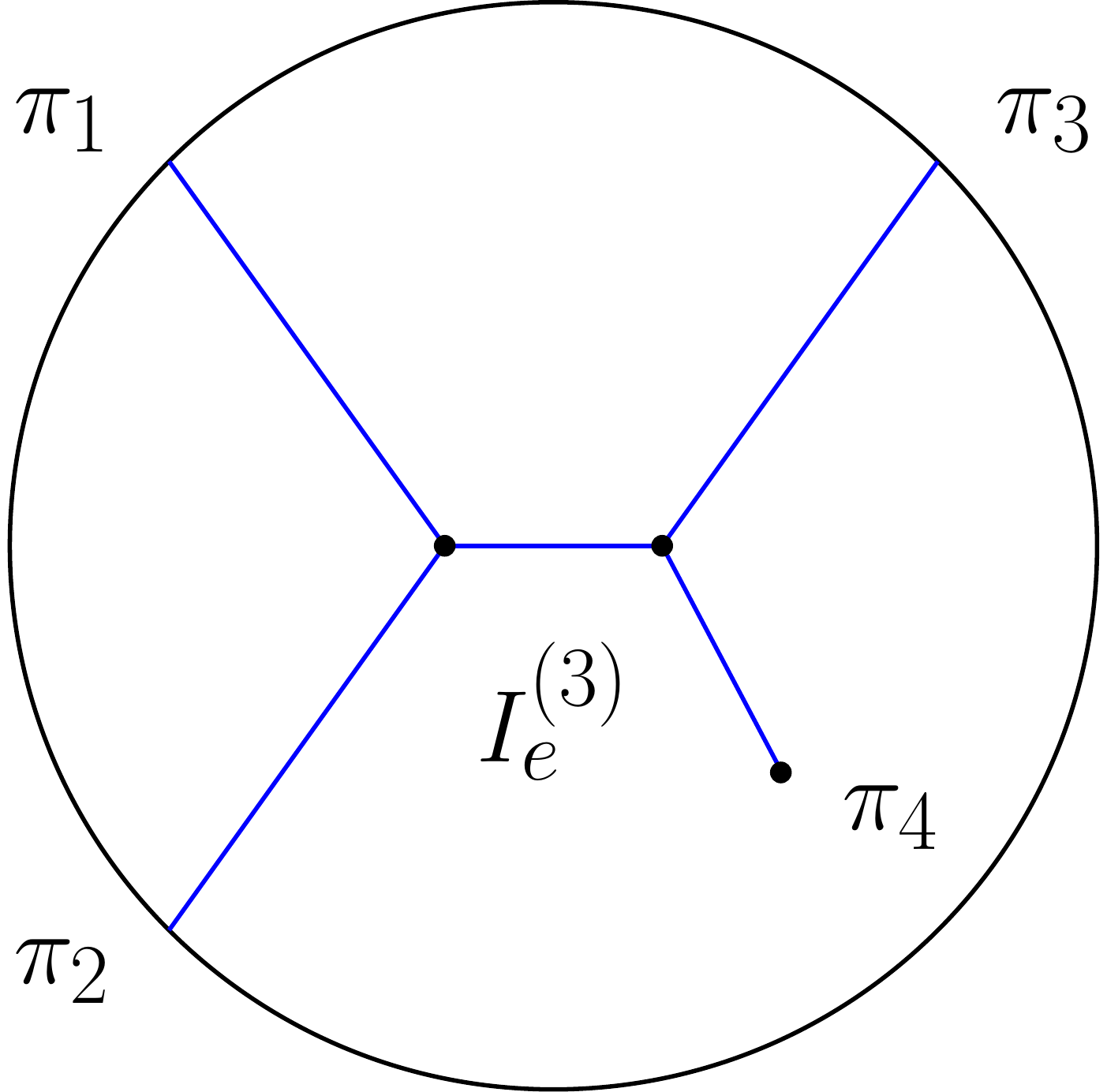}
\caption{Four-point vertex (left) and exchange (right) Witten diagrams used to compute the one-point function of the axial current. Lines ending at the boundary represent bulk-to-boundary propagators and introduce factors proportional to the source, while lines connecting points in the interior represent bulk propagators.}\label{fig:wittendiag}
\end{center}
\end{figure}

The inhomogeneous terms at $O(\epsilon^3)$ can be split in two types of contributions, corresponding to different Witten diagrams. One contribution, $I^{(3)}_v$, corresponds to a four-point vertex, a diagram where four gauge field propagators join at a single point in the bulk. The other contribution, $I^{(3)}_e$, takes the form of an exchange diagram between two three-point vertex, the field propagates in the bulk between two points and there are two other propagators at each point. We have sketched the two Witten diagrams in Fig.~\ref{fig:wittendiag}. For each type we are interested only in terms with massless poles that will be the only ones contributing to the pion scattering amplitude. The leading pole contribution to the scattering amplitude has a massless pole for each external leg. Within $O(\epsilon^3)$, the only leading pole contributions are those terms with two (three) massless pole factors in $I_Z$ ($I_\mu$). From \eqref{eq:order1sol} one can see that the only terms depending on the first order Abelian field strength that contain massless poles are those proportional to $A_Z^{(1)\, a}$ or $f_{Z\mu}^{(1)\,a}$. This fact will significantly reduce the number of terms we need to consider.

The details of the calculation have been relegated to the Appendix~\ref{app:calculation}. We identify three types of terms that can give a contribution to the four-point function:
\begin{itemize}

\item {\em Contributions from $O(F^4)$ terms in the canonical momentum.} The $O(\epsilon)$ solution to the gauge fields could give a direct contribution to the four-point function of the axial current through the $\sim F^3$ term in the canonical momentum, $\Pi^{[4]}$ in \eqref{eq:canmom}. However, it turns out that this does not give any contribution to the leading pole, so we can ignore it for the purpose of computing the scattering amplitude.

\item {\em Contributions from $O(F^2)$ terms.} The non-Abelian terms in the field strength introduce a cubic coupling among gauge fields in the bulk. Joining two such vertices with a bulk gauge field propagator results in an exchange Witten diagram (cf. right panel of Fig.~\ref{fig:wittendiag}) that does give a contribution to the leading pole of the four-point function. The contribution is:
\begin{equation}\label{eq:exchange4p}
\begin{split}
&\vev{J_5^{\mu_1\,a_1}(p_1)J_5^{\mu_2\,a_2}(p_2)J_5^{\mu_3\,a_3}(p_3)J_5^{\mu_4\,a_4}(p_4)}_e  \\ 
&\simeq  -2 i \f^2\left(\prod_{i=1}^4\frac{p_i^{\mu_i}}{p_i^2}\right) \delta_{\sum_{i=1}^4 p_i} \Big[\Big\{(p_1\cdot p_2)-\frac{4b}{\pi^3 M_{\text{KK}}^2} \left[(p_1\cdot p_3)^2+(p_1\cdot p_4)^2-2(p_1\cdot p_2)^2 \right] \Big\}\delta^{a_1 a_2}\delta^{a_3 a_4}\\
&+(2\leftrightarrow 3)+(2\leftrightarrow 4)\Big] \ ,
\end{split}
\end{equation}
where $b$ is given in (\ref{eq:coefsSS}). Non-Abelian terms also introduce quartic couplings among gauge fields, but these do not contribute to the leading pole.

\item {\em Contributions from $O(F^4)$ terms in the D$8$-brane action.} These terms introduce a quartic coupling between the Abelianized  field strengths. This quartic coupling results in a vertex Witten diagram (cf. left panel of Fig.~\ref{fig:wittendiag}) that also contributes to the leading pole of the four-point function as follows:
\begin{equation}\label{eq:vertex4p}
\begin{split}
&\vev{J_5^{\mu_1\,a_1}(p_1)J_5^{\mu_2\,a_2}(p_2)J_5^{\mu_3\,a_3}(p_3)J_5^{\mu_4\,a_4}(p_4)}_v \\ 
&\simeq   i \frac{\f^2}{M_{\text{KK}}^2}\frac{3^5\Gamma\left(\frac{13}{6}\right)}{4\sqrt{\pi} \Gamma\left(\frac{8}{3}\right)\lambda_{\text{YM}}^2} \left(\prod_{i=1}^4\frac{p_i^{\mu_i}}{p_i^2}\right) \delta_{\sum_{i=1}^4 p_i}\left(\delta^{a_1 a_2}\delta^{a_3 a_4}+\delta^{a_1 a_3}\delta^{a_2 a_4}+\delta^{a_1 a_4}\delta^{a_2 a_3}\right)\\ 
& \qquad\qquad\times \left[(p_1\cdot p_2)(p_3\cdot p_4)+(2\leftrightarrow 3)+(2\leftrightarrow 4) \right] \ .
\end{split}
\end{equation}
\end{itemize}

%%%%%%%%%%%%%%%%%%%%%%%%%%%%%%%%%%%%%%%%%%%%%
\subsection{Scattering amplitude}
%%%%%%%%%%%%%%%%%%%%%%%%%%%%%%%%%%%%%%%%%%%%%

We are now ready to extract the pion correlators and the scattering amplitude. Recall the expression in \eqref{eq:axial4p}, where the current four-point function is given in terms of the pion four-point function. We can thus read off the pion four-point function from \eqref{eq:exchange4p} and \eqref{eq:vertex4p} by removing the $p_i^{\mu_i}$ factors and dividing by $\f^4$. Furthermore, the scattering amplitude appears in the four-point function of the pions, recall \eqref{eq:pion4p}, as the residue of the leading pole, once the delta function corresponding to the momentum conservation and an $i$ factor have been factored out.

The resulting amplitude has the expected structure \eqref{eq:pionamplitude}, with the exchange contribution \eqref{eq:exchange4p} being
\bea
 A_e(s,t,u) & = & -\frac{2(p_1\cdot p_2)}{\f^2}+\frac{8b}{\pi^3\f^2  M_{\text{KK}}^2} \left[(p_1\cdot p_3)^2+(p_1\cdot p_4)^2-2(p_1\cdot p_2)^2 \right] \nonumber \\
& = & \frac{s}{\f^2}+\frac{2b}{\pi^3\f^2  M_{\text{KK}}^2} \left[t^2+u^2-2s^2\right] \ .
\eea
The first term in the exchange contribution agrees with the first term in Weinberg's amplitude \eqref{eq:weinbergA}. This therefore proves that the dual holographic derivation is indeed consistently capturing pion dynamics from spontaneous chiral symmetry breaking. The remaining terms take the form expected from vector meson exchange, giving a contribution to the $O(p^4)$ terms which agrees with the expressions in \eqref{eq:coefsSS},
\begin{equation}
L_2^e=-L_1^e=\frac{\Nc\lambda_{\text{YM}}}{ 6^3 \pi^7}b\ .
\end{equation}

The vertex contribution \eqref{eq:vertex4p} contains further terms that we can associate to pion self-interactions
\bea
 A_v(s,t,u) & = & \frac{1}{\f^2 M_{\text{KK}}^2}\frac{3^5\Gamma\left(\frac{13}{6}\right)}{4\sqrt{\pi} \Gamma\left(\frac{8}{3}\right)\lambda_{\text{YM}}^2} \left[(p_1\cdot p_2)(p_3\cdot p_4)+(2\leftrightarrow 3)+(2\leftrightarrow 4)  \right] \nonumber\\
 &= & \frac{3^5\Gamma\left(\frac{13}{6}\right)}{2^4\sqrt{\pi} \Gamma\left(\frac{8}{3}\right)\lambda_{\text{YM}}^2}  \frac{s^2+t^2+u^2}{\f^2 M_{\text{KK}}^2} \ .
\eea
Comparing with Weinberg's amplitude \eqref{eq:weinbergA}, and using the expression for $\f^2$ in \eqref{eq:coefsSS}, we can read off an additional contribution to the value of the coefficients in the pion effective action
\begin{equation}\label{eq:LECresult}
 L_2^v=2 L_1^v=\frac{18 \Nc}{(4\pi)^4\lambda_{\text{YM}}} \frac{\Gamma\left(\frac{13}{6}\right)}{\sqrt{\pi}\Gamma\left(\frac{8}{3}\right)} \ . 
\end{equation}
The numerical value of the constant factor with the gamma functions is approximately $\frac{\Gamma\left(\frac{13}{6}\right)}{\sqrt{\pi}\Gamma\left(\frac{8}{3}\right)}\approx 0.406$. 

The full value of the LECs when the vector bosons are integrated out is the sum of the exchange and vertex contributions $\hat{L}_1=L_1^e+L_1^v$, $\hat{L}_2=L_2^e+L_2^v$. However, if the vector bosons are kept in the effective action, then only the vertex contributions produce non-vanishing LECs in the chiral Lagrangian $L_1=L_1^v$, $L_2=L_2^v$.

%%%%%%%%%%%%%%%%%%%%%%%%%%%%%%%%%%%%%%%%%%%%%
\subsection{Integrating out vector bosons and Hidden Local Symmetry}
%%%%%%%%%%%%%%%%%%%%%%%%%%%%%%%%%%%%%%%%%%%%%

One might find it convenient to integrate out only the vector mesons above some mass threshold, in particular keeping only the lightest vector mode. In the scattering calculation the contribution from each massive vector mode can be identified using an eigenfunction expansion of the bulk propagator \eqref{eq:Greenfunc} entering in the exchange diagram
\begin{equation}
 G(Z,Z_1;q)=-\sum_{n=1}^\infty \frac{\psi_n(Z) \psi_n(Z_1)}{q^2+m_n^2} \ ,
\end{equation}
where the eigenfunctions satisfy the following equations
\begin{equation}
\begin{split}
&\partial_Z\left( u(Z)^3 \partial_Z \psi_n(Z)\right)+\frac{m_n^2}{u(Z)}\psi_n(Z)=0 \ , \ \lim_{Z\to\pm \infty}\psi_n(Z)=0 \\
&\int_{-\infty}^\infty dZ \frac{\psi_n(Z)\psi_m(Z)}{u(Z)}=\delta_{nm} \ .
\end{split}
\end{equation}
It should be noted that the $O(p^2)$ contribution to the pion scattering amplitude is obtained from the $q^2=0$ value of the bulk propagator. When expressed in this form, the value of $\f$ obtained from the amplitude is determined by the exchange of an infinite tower of massive modes. Notice that a bulk vertex diagram or a bulk exchange diagram might not correspond necessarily to vertex or exchange processes in the field theory dual, although it seems natural to do this identification. Under this assumption, though, the vector bosons corresponding to mass eigenstates would not couple to the pions as Weinberg's $\rho$ meson discussed in Sec.~\ref{sec:weinbergact} and the effective action of the pion field (before integrating out the massive modes) do not have the $O(\partial^2 \pi^4)$ terms expected in the chiral Lagrangian. Then, the effective action written in terms of these fields would not comply with HLS invariance in any obvious way.

The LECs obtained from integrating out all massive vector modes except the lightest one would na\"ively be obtained by replacing the full bulk propagator by the truncated sum 
\begin{equation}
\begin{split}
G_{n>1}(Z,Z_1;q)=-\sum_{n=2}^\infty \frac{\psi_n(Z) \psi_n(Z_1)}{q^2+m_n^2} \ .
\end{split}
\end{equation}
In this case the effective action of the pion would have the expected $O(\partial^2\pi^4)$ terms of the chiral Lagrangian, with an effective value of $\f$ determined by the modes that have been integrated out. However, the right value of $\f$ measured in the full scattering amplitude would be recovered only after considering the tree level exchange by the lightest vector meson. 

The issue with HLS invariance of mass eigenstates has been pointed out for instance in \cite{Harada:2006di,Harada:2010cn,Harada:2014lza}, where an alternative basis of radial functions has been proposed to construct an explicitly HLS invariant action. It would be interesting if the scattering amplitude calculation could be connected to the HLS covariant formalism in some way.

%%%%%%%%%%%%%%%%%%%%%%%%%%%%%%%%%%%%%%%%%%%%%
%%%%%%%%%%%%%%%%%%%%%%%%%%%%%%%%%%%%%%%%%%%%%
\section{Discussion}\label{sec:discussion}
%%%%%%%%%%%%%%%%%%%%%%%%%%%%%%%%%%%%%%%%%%%%%
%%%%%%%%%%%%%%%%%%%%%%%%%%%%%%%%%%%%%%%%%%%%%

In this paper we presented a computation of the pion scattering amplitude for two massless flavors in the WSS model \cite{Witten:1998zw,Sakai:2004cn,Sakai:2005yt}. Our main result is given in \eqref{eq:LECresult}. 
These would be the coefficients of pion self-interactions in the effective action before vector mesons have been integrated out. It differs from the result that was extracted from the effective action \eqref{eq:actionWSS} appearing in the seminal work \cite{Sakai:2004cn,Sakai:2005yt}, where the coefficients can be removed by a field redefinition of the vector meson fields that puts the action in the standard form \eqref{eq:actionWSS2}.
The main qualitative difference between the previously quoted results and the actual value of the LECs can be summarized in the following table
\begin{center}
\begin{tabular}{c|c}
previous result & scattering result \\ \hline
$L_2^{SU(2)}=-L_1^{SU(2)}$ & $L_2^{SU(2)}=2L_1^{SU(2)}$\\
$L_i \sim \Nc\lambda_{YM}$ & $L_i\sim \Nc\lambda_{YM}^{-1}$
\end{tabular} 
\end{center}
These relations hold before massive vector mesons, corresponding to mass eigenstates in the gravity dual, have been integrated out. At lower energies, vector meson exchange contributions modify the LECs, and we find that the relation $\hat{L}_2^{SU(2)}\simeq -\hat{L}_1^{SU(2)}$, with the values that were originally proposed, holds up to the $1/\lambda_{\text{YM}}^2$ corrections in the strong coupling limit that we have computed.\footnote{The modified LECs are measured always at energies much below the vector meson masses, where the vector meson exchange contribution can be approximated by an effective local pion self-interaction, as in \eqref{eq:effectiveLEC}.} In other cases where the massive vector bosons do not correspond to the mass eigenstates of the holographic dual (see, {\emph{e.g.}}, \cite{Harada:2006di,Harada:2010cn,Harada:2014lza}), the identification of the $L_i$ coefficients is not as straightforward, but there will be a correction to the coefficients of the effective action $\sim 1/\lambda_{\text{YM}}^2$ such that the low momentum amplitude reproduces our results. 

We will elaborate more on the implications below. In addition to the WSS model, the relation $L_2^{SU(2)}=-L_1^{SU(2)}$ was also obtained in bottom-up models that followed essentially the same approach to derive the effective action \cite{DaRold:2005mxj,Hirn:2005nr,Panico:2007qd,Domenech:2010aq,Colangelo:2012ipa,Domokos:2014ura}, with slight deviations from the classical value when other fields are integrated out \cite{DaRold:2005vr}. The relation between $L_1$ and $L_2$ could possibly be modified already at the classical level if the LECs are extracted from the pion scattering amplitude following the method used in our work, depending on whether they are completely determined by vector meson exchange or not. However, in the bottom-up models there is an additional bilinear field dual to the chiral condensate whose effect in the scattering amplitude should be studied more carefully, so we cannot extrapolate directly the WSS results to those models. We also expect that the value of the LECs in the effective action depend generically on which fields have been integrated out.

The result for two flavors determines already the four-derivative terms of the chiral effective action for an arbitrary number of flavors $\Nf$. The reason is that, in the large-$\Nc$ limit, the only contributions to the chiral effective action with $O(\Nc)$ coefficients are single trace terms (see, {\emph{e.g.}}, \cite{Manohar:1998xv})
\begin{equation}
 {\cal L}_{p^4}=L_3^{SU(\Nf)} \textrm{Tr}\left(\partial_\mu\Sigma^{\dagger}\partial^\mu\Sigma\partial_\nu\Sigma^{\dagger}\partial^\nu\Sigma\right)+\tilde{L}_3^{SU(\Nf)}\textrm{Tr}\left(\partial_\mu\Sigma^{\dagger}\partial_\nu\Sigma\partial^\mu\Sigma^{\dagger}\partial^\nu\Sigma\right) \ .
\end{equation}
But for $\Nf=3$ the last term can be rewritten as combination of the term proportional to $L_3^{SU(\Nf)}$ and double-trace terms, giving
\begin{equation}
 L_1^{SU(3)}=\frac{1}{2}\tilde{L}_3^{SU(\Nf)}\ ,  \ L_2^{SU(3)}=\tilde{L}_3^{SU(\Nf)} \ , \ L_3^{SU(3)}=L_3^{SU(\Nf)}-2 \tilde{L}_3^{SU(\Nf)} \ .
\end{equation}
Then, for $\Nf=2$, the two independent terms that are left have coefficients
\begin{equation}
 L_1^{SU(2)}=\frac{L_3^{SU(\Nf)}-\tilde{L}_3^{SU(\Nf)}}{2}\ , \ L_2^{SU(2)}=\tilde{L}_3^{SU(\Nf)} \ .
\end{equation}
Therefore for an arbitrary number of flavors we claim that the result is simply $\tilde{L}_3^{SU(\Nf)}=2L_3^{SU(\Nf)}=L_2^{SU(2)}$ with $L_2^{SU(2)}$ determined by the equation \eqref{eq:LECresult}.

We have limited our calculation to the leading contributions to the pion scattering amplitude at low momentum. Higher order corrections in momentum can be computed systematically using the expressions for the bulk propagator introduced in \eqref{eq:Gqexp} and \eqref{eq:Gqexpcoef}. For momentum of the order or larger than the confinement scale the full form of the bulk propagator would be necessary. At very large momentum, though, stringy $\alpha'$ corrections become relevant, and the scattering amplitude might be approximated by integrating a flat space string amplitude along the holographic coordinate \cite{Polchinski:2001tt,Polchinski:2002jw,Brower:2006ea,Domokos:2009hm,Liu:2022out,Amorim:2021gat}, an approach that has been applied in \cite{Bianchi:2021sug} to holographic duals of confining theories. However, as discussed there, in the high energy regime the WSS model is dual to a six-dimensional theory, so not suitable for a comparison with high-momentum scattering in QCD.

A further important extension of our work would be to study pion scattering with nonzero quark masses. Within the WSS model this requires considering non-antipodal embeddings and in the presence of an additional ``tachyon'' field \cite{Bergman:2007pm} that is dual to a quark bilinear. In bottom-up models, such as the original AdS/QCD hard wall model \cite{Erlich:2005qh}, or V-QCD \cite{Jarvinen:2011qe}, the tachyon field is already included, and in addition the V-QCD model also has quartic terms in the action of the gauge fields dual to the flavor currents. The quark mass explicitly breaks the axial flavor symmetry and gives a mass to the pions. ChPT can still be used, but with additional terms in the effective action. 

When quarks are massive, there is a finite scattering length determined by the pion mass that can be obtained from the pion scattering amplitude at leading order in low momentum (see, {\emph{e.g.}}, \cite{Kaplan:2005es}). A general prescription and a few examples to compute the scattering length of massive particles in strongly coupled theories using the holographic dual were given in \cite{Hoyos:2019kzt,Hoyos:2020fjx}. Higher momentum corrections to the scattering amplitude can also be computed and used to constrain the LECs appearing in the chiral effective action. It would be interesting to extract the scattering amplitude in holographic models and compare with recent large-$\Nc$ lattice results \cite{Baeza-Ballesteros:2022azb}.

Finally, let us pronounce the main message of our work. That is, the calculation of scattering amplitudes should be applied also to other setups. As our example on the WSS model shows, this might be a requisite step to correctly identify the low energy effective theory that captures the dynamics encoded by the holographic dual.

%%%%%%%%%%%%%%%%%%%%%%%%%%%
\section*{Acknowledgements}
%%%%%%%%%%%%%%%%%%%%%%%%%%%

We would like to thank M.~J\"arvinen, T.~Sakai, and S.~Sugimoto for careful readings of the manuscript and for many useful comments. We thank U.~G\"ursoy for useful comments motivating us to look further into the vector boson exchange contributions. C.~H. is partially supported by the AEI through the Spanish grant PGC2018-096894-B-100 and by FICYT through the Asturian grant SV-PA-21-AYUD/2021/52177. N.~J. is supported in part by the Academy of Finland grant no. 1322307.

\appendix

%%%%%%%%%%%%%%%%%%%%%%%%%%%%%%%%%%%%%%%%%%%%%
%%%%%%%%%%%%%%%%%%%%%%%%%%%%%%%%%%%%%%%%%%%%%
\section{Calculation of the four-point function}\label{app:calculation}
%%%%%%%%%%%%%%%%%%%%%%%%%%%%%%%%%%%%%%%%%%%%%
%%%%%%%%%%%%%%%%%%%%%%%%%%%%%%%%%%%%%%%%%%%%%

In this Appendix we spell out the details of the calculation of the axial current four-point function following the procedure described in the main text. As listed in section \ref{sec:fourpoint}, there are three possible contributions we have to study: $O(F^4)$ terms in the action and the associated terms in the canonical momentum as well as $O(F^2)$ terms in the action.

%%%%%%%%%%%%%%%%%%%%%%%%%%%%%%%%%
\subsection{Contributions from $O(F^4)$ terms in the canonical momentum: contact terms}
%%%%%%%%%%%%%%%%%%%%%%%%%%%%%%%%%

The $O(\epsilon)$ solution to the gauge fields could give a direct contribution to the four-point function of the axial current through the $\sim F^3$ term in the canonical momentum, $\Pi^{[4]}$ in \eqref{eq:canmom}. The leading pole contribution involves just the $f_{Z\mu}^{(1)}$ components
\bea
\Pi^{[4]\,\mu}_a & \simeq & \frac{1}{6}\left[4f_{Z\nu }^b\left(f^{a\,\mu Z}f^{b\,Z \nu}+f^{b\,\mu Z}f^{a\,Z \nu}\right)+4f_{\,Z\nu}^a f^{b\,\mu Z}f^{b\,Z \nu}\right.   \nonumber\\
& & \left.+2\left(f^{a\,Z\mu}f_{Z\nu}^b f^{b\,Z\nu}+2f^{b\,Z\mu}f^{b\,Z\nu}f_{\,Z\nu}^a\right)\right] \nonumber\\
 & = & -\frac{1}{3}\left[f_{Z\nu}^b f^{b\,Z\nu} f^{a\,Z\mu}+2f_{Z\nu}^b f^{a\,Z\nu} f^{b\,Z\mu}\right] \ .\label{eq:Pi4}
\eea
However, this contribution to the canonical momentum vanishes when $|Z|\to \infty$, since from \eqref{eq:fZ1},
\begin{equation}
\pi_a^\mu \sim u^2 \Pi_a^{[4]\,\mu } \sim u^4 \phi_0(Z)^3 \underset{|Z|\to \infty}{\sim} \frac{1}{|Z|^{10/3}} \longrightarrow 0 \ . 
\end{equation}
Therefore, there is no contribution to the expectation value of the axial current or the four-point function from these terms.

%%%%%%%%%%%%%%%%%%%%%%%%%%%%%%%%%
\subsection{Contributions from $O(F^2)$ terms: exchange diagram}
%%%%%%%%%%%%%%%%%%%%%%%%%%%%%%%%%

Let us discuss the contributions coming from terms quadratic in the field strength. The $O(A^4)$ non-Abelian quartic term in the action introduces $O(\epsilon^3)$ non-linear terms in the equations of motion, proportional to
\begin{equation}
\begin{split}
&I_Z^{[2]\,a}\sim \epsilon^{abc}\epsilon^{dec} \eta^{\mu\nu} A_\mu^{(1)\,b} A_\nu^{(1)\,d} A_Z^{(1)\,e} \\
&I_\mu^{[2]\,a}\sim \epsilon^{abc}\epsilon^{dec}  A_Z^{(1)\,b} A_Z^{(1)\,d} A_\mu^{(1)\,e} \ , \ \epsilon^{abc}\epsilon^{dec}  \eta^{\alpha\beta} A_\alpha^{(1)\,b} A_\beta^{(1)\,d} A_\mu^{(1)\,e}.
\end{split}
\end{equation}
The antisymmetry of the structure constants guarantees that there are no $\sim (A_Z^{(1)})^3$ terms. But these would be the only terms contributing to the leading pole. As they are absent, we can neglect the contributions coming from the quartic terms in the gauge potentials.

The non-linear terms in the equations of the bulk gauge field originating from the $O(A^3)$ terms in the action are
\begin{equation}
\begin{split}
&I_Z^{[2]\,a}=-u^3\epsilon^{abc}\left[ \eta^{\alpha\beta}\partial_\alpha\left(A_\beta^b A_Z^c\right)+\eta^{\alpha\beta} A_\alpha^b F_{\beta Z}^c \right] \\
&I_\mu^{[2]\,a}=-\epsilon^{abc}\left[\partial_Z\left(u^3 A_Z^b A_\mu^c\right)+u^3 A_Z^b F_{Z\mu}^c+ \frac{1}{u}\eta^{\alpha\beta}\partial_\alpha\left(A_\beta^b A_\mu^c\right)+\frac{1}{u}\eta^{\alpha\beta} A_\alpha^b F_{\beta\mu}^c\right] \ .
\end{split}
\end{equation}
Let us first identify the vertex contributions, they are those with three field factors
\begin{equation}
\begin{split}
&I_{Z\,v}^{[2]\,(3)\,a}=-u^3\epsilon^{abc}\epsilon^{cde}\eta^{\alpha\beta} A_\alpha^{(1)\,b} A_\beta^{(1)\,d} A_Z^{(1)\,e} \\
&I_{\mu\,v}^{[2]\,(3)\, a}=-\epsilon^{abc}\epsilon^{cde}\left[ u^3 A_Z^{(1)\,b} A_Z^{(1)\,d} A_\mu^{(1)\,e}+\frac{1}{u}\eta^{\alpha\beta} A_\alpha^{(1)\,b} A_\beta^{(1)\,d} A_\mu^{(1)\,e}\right] \ .
\end{split}
\end{equation}
Since all the terms have less massless pole factors than those required to give a contribution to the leading pole term, we can neglect these contributions in the correlator.

Let us now move on to the exchange contributions, they are those with two field factors
\bea
 I_{Z\, e}^{[2]\,(3)\, a} & = & -u^3\epsilon^{abc}\left[ \eta^{\alpha\beta}\partial_\alpha\left(A_\beta^b A_Z^c\right)+\eta^{\alpha\beta} A_\alpha^b f_{\beta Z}^c \right]^{(1),(2)} \\
 I_{\mu\,e}^{[2]\,(3)\, a} & = & -\epsilon^{abc}\left[\partial_Z\left(u^3 A_Z^b A_\mu^c\right)+u^3 A_Z^b f_{Z\mu}^c+ \frac{1}{u}\eta^{\alpha\beta}\partial_\alpha\left(A_\beta^b A_\mu^c\right)+\frac{1}{u}\eta^{\alpha\beta} A_\alpha^b f_{\beta\mu}^c\right]^{(1),(2)} \ .
\eea
The superscript notation means that from the two factors in each term in the brackets, one should be $O(\epsilon)$ and the other $O(\epsilon^2)$, and we must consider all possibilities. In order to compute this contribution we will need the $O(\epsilon^2)$ inhomogeneous terms as well
\bea
 I_Z^{[2]\,(2)\, a} & = & -u^3\epsilon^{abc}\left[ \eta^{\alpha\beta}\partial_\alpha\left(A_\beta^{(1)\,b} A_Z^{(1)\,c}\right)+\eta^{\alpha\beta} A_\alpha^{(1)\,b} f_{\beta Z}^{(1)\,c} \right] \\
 I_\mu^{[2]\,(2)\, a} & = & -\epsilon^{abc}\left[\partial_Z\left(u^3 A_Z^{(1)\,b} A_\mu^{(1)\,c}\right)+u^3 A_Z^{(1)\,b} f_{Z\mu}^{(1)\,c}+ \frac{1}{u}\eta^{\alpha\beta}\partial_\alpha\left(A_\beta^{(1)\,b} A_\mu^{(1)\,c}\right)+\frac{1}{u}\eta^{\alpha\beta} A_\alpha^{[2]\,(1)\,b} f_{\beta\mu}^{(1)\,c}\right] \ .
\eea
In this case there is a contribution to the leading pole term of the correlator from terms in $I_M^{(2)}$ with two massless pole factors, corresponding to two external legs of the exchange Witten diagram joining in a vertex with the internal leg. This leaves only one term that needs to be considered
\bea
 I_Z^{[2]\,(2)\, a}& \simeq & 0 \\
 I_\mu^{[2]\,(2)\, a} & \simeq & -\epsilon^{abc}u^3 A_Z^{(1)\,b} f_{Z\mu}^{(1)\,c} \ .
\eea
Taking this into account, we can set $A_Z^{(2)\,a}\simeq 0$ to compute the leading pole contribution. This leaves
\bea
 I_{Z}^{[2]\,(3)\, a} & \simeq & -u^3\epsilon^{abc}\left[ \eta^{\alpha\beta}\partial_\alpha\left(A_\beta^{(2)\,b} A_Z^{(1)\,c}\right)+\eta^{\alpha\beta} A_\alpha^{(2)\,b} f_{\beta Z}^{(1)\,c} \right] \\
 I_{\mu}^{[2]\,(3)\, a} & = & -\epsilon^{abc}\left[\partial_Z\left(u^3 A_Z^{(1)\,b} A_\mu^{(2)\,c}\right)+u^3 A_Z^{(1)\,b} f_{Z\mu}^{(2)\,c}\right] \ .
\eea

Finally, there could had been an $O(\epsilon^3)$ exchange contribution where the internal leg of the Witten diagram is the Abelian component of the D$8$-brane gauge field and the vertices are determined by the Wess-Zumino action \eqref{eq:WZlag}. To compute this contribution one should first find the $O(\epsilon^2)$ solution for the Abelian field. The inhomogeneous terms in the Abelian field equation are proportional to
\begin{equation}
 I_Z^{[WZ]}\sim \epsilon^{Z\mu\nu\alpha\beta}f_{\mu\nu}^{(1)\,a}f_{\alpha\beta}^{(1)\,a}\ , \ I_\mu^{[WZ]}\sim \epsilon_\mu^{\ \nu Z\alpha\beta} f_{\nu Z}^{(1)\,a} f_{\alpha\beta}^{(1)\,a}\ .
\end{equation}
But none of these terms has two massless pole factors, so they do not contribute to the leading pole term in the correlator of the axial current.

Moving on to the calculation of the solution to the gauge field, given the $O(\epsilon)$ solutions  \eqref{eq:order1sol} with momenta $p_i$, $p_j$, the $O(\epsilon^2)$ inhomogeneous term contributing to the leading pole would be
\begin{equation}
I_\mu^{[2]\,(2)\,a}(p_i,p_j;Z,q)\simeq \frac{4}{\pi}\phi_0(Z) \int_{p_i}\int_{p_j} \delta_{q-p_i-p_j}\, \frac{ip_i^\alpha}{p_i^2}\left(\delta_\mu^{\ \beta}-\frac{p_{j\,\mu} p_j^\beta}{p_j^2} \right)\epsilon^{abc}\hat{A}_{5\,\alpha}^b(p_i)\hat{A}_{5\,\beta}^c(p_j) \ ,
\end{equation}
where we are using as shorthand notation
\begin{equation}
 \int_p\equiv \int \frac{d^4 p}{(2\pi)^4}\, \ \ \delta_p=(2\pi)^4\delta^{(4)}(p) \ .
\end{equation}
One can check using the symmetries of the integrand that $q^\mu I_\mu^{(2)}=0$, so this is a transverse term and in addition it is independent of the radial coordinate $Z$. We can further simplify this expression by keeping only the leading pole term
\begin{equation}
 I_\mu^{[2]\,(2)\,a}(p_i,p_j;Z,q)\simeq i_\mu^{(2)\,a}(p_i,p_j;q)\phi_0(Z) \ ,
\end{equation}
where, in order to make expressions more manageable we have defined
\begin{equation}\label{eq:i2mu}
 i_\mu^{(2)\,a}(p_i,p_j;q)=-\frac{4}{\pi} \int_{p_i}\int_{p_j} \delta_{q-p_i-p_j}\, \frac{ip_i^\alpha}{p_i^2}\frac{p_{j\,\mu} p_j^\beta}{p_j^2}\epsilon^{abc}\hat{A}_{5\,\alpha}^b(p_i)\hat{A}_{5\,\beta}^c(p_j) \ .
\end{equation}
Then, from \eqref{eq:AparSol}, the $O(\epsilon^2)$ gauge field solution is 
\begin{equation}
 A_\mu^{(2)\,a}(Z,q)\simeq \int dZ_1 G(Z,Z_1;q^2)I_\mu^{[2]\,(2)\,a}(p_i,p_j;Z_1,q)\simeq i_\mu^{(2)\,a}(p_i,p_j;q)\left[\Phi^{(2)}(Z)+q^2 \widetilde{\Phi}^{(2)}(Z)\right]+O(q^4)  \ ,
\end{equation}
where
\begin{equation}
\begin{split}
& \Phi^{(2)}(Z)=\int dZ_1 G(Z,Z_1;0)\phi_0(Z_1)=\frac{\pi}{2}\psi_0(Z)(\psi_0(Z)-1) \\
& \widetilde{\Phi}^{(2)}(Z)=\int dZ_1 G^{(1)}(Z,Z_1)\phi_0(Z_1)\ .
\end{split}
\end{equation}
Here we are introducing an additional approximation, not only $p_i^2\simeq 0$, $p_j^2\simeq 0$ are close to lightlike values, but also we assume $|(p_i+p_j)^2|\ll 1$, i.e., low energy and momentum for the external pions.

Next, we compute the $O(\epsilon^3)$ inhomogeneous terms
\bea
 I_{Z}^{[2]\,(3)\, a} & \simeq & -2u^3\epsilon^{abc}\eta^{\alpha\beta} A_\alpha^{(2)\,b} \partial_\beta A_Z^{(1)\,c} \\
 I_{\mu}^{[2]\,(3)\, a} & \simeq & -2\epsilon^{abc} u^3 A_Z^{(1)\,b}\partial_Z A_\mu^{(2)\,c} \ ,
\eea
where we have used $\partial_Z(u^3 A_Z^{(1)})=0$, $\eta^{\alpha\beta}\partial_\alpha A_\beta^{(2)\,a}=0$ and kept the leading pole terms only. Assigning momentum $p_k$ to the $O(\epsilon)$ factors
\bea
 I_{Z}^{[2]\,(3)\, a}(Z,p_l) & \simeq & -\frac{4}{\pi} \left[\Phi^{(2)}(Z)+q^2\widetilde{\Phi}^{(2)}(Z)\right]\int_{p_k} \int_q \delta_{p_l-p_k-q} \epsilon^{abc}i_\alpha^{(2)\,b}(p_i,p_j;q)\frac{p_k^\alpha p_k^\beta}{p_k^2}\hat{A}_{5\,\beta}^c(p_k) \\
 I_{\mu}^{[2]\,(3)\, a} & \simeq & -\frac{4}{\pi} \left[\partial_Z\Phi^{(2)}(Z)+q^2 \partial_Z\widetilde{\Phi}^{(2)}(Z) \right]\int_{p_k} \int_q \delta_{p_l-p_k-q}  \epsilon^{abc}i_\mu^{(2)\,b}(p_i,p_j;q)\frac{ip_k^\alpha}{p_k^2}\hat{A}_{5\,\alpha}^c(p_k) \ .
\eea
Let us define
\begin{equation}\label{eq:i3mu}
 i_\mu^{(3)\,a}(p_i,p_j,p_k,p_l)= -\frac{4}{\pi}  \int_{p_k} \int_q \delta_{p_l-p_k-q}  \epsilon^{abc}i_\mu^{(2)\,b}(p_i,p_j;q)\frac{ip_k^\alpha}{p_k^2}\hat{A}_{5\,\alpha}^c(p_k) \ ,
\end{equation}
then
\bea
 I_{Z}^{[2]\,(3)\, a}(Z,p_l) & \simeq & -\left[\Phi^{(2)}(Z)+q^2 \widetilde{\Phi}^{(2)}(Z) \right] i p_l^\mu i_\mu^{(3)\,a}(p_i,p_j,p_k,p_l) \\
 I_{\mu}^{[2]\,(3)\, a} & \simeq & \left[{\Phi^{(2)}}'(Z)+q^2 {\widetilde \Phi}^{(2)}{}'(Z)\right] i_\mu^{(3)\,a}(p_i,p_j,p_k,p_l) \ .
\eea
Following \eqref{eq:AparSol} and \eqref{eq:AperSol}, the $O(\epsilon^3)$ solution for the gauge potential is
\begin{equation}
\begin{split}
 &A_\mu^{[2]\,(3)\,a}(Z,p_l)\simeq \left(\delta_\mu^{\ \nu}-\frac{p_{l\,\mu}p_l^\alpha}{p_l^2} \right)i_\alpha^{(3) \,a}(p_i,p_j,p_k,p_l)\left[\Phi^{(3)}(Z) +q^2 {\widetilde \Phi}^{(3)}(Z)\right]  \\
 &A_Z^{[2]\,(3)\,a}(Z,p_l)=\frac{\Phi^{(2)}(Z)+q^2{\widetilde \Phi}^{(2)}(Z)}{1+Z^2}\frac{i p_l^\alpha}{p_l^2} i_\alpha^{(3)\,a}(p_i,p_j,p_k,p_l) \ ,
\end{split}
\end{equation}
where
\begin{equation}\label{eq:phi3}
\begin{split}
& \Phi^{(3)}(Z)=\int dZ_1 G(Z,Z_1;0){\Phi^{(2)}}'(Z_1)=\frac{\pi}{6} \arctan(Z)\psi_0(Z)(\psi_0(Z)-1) \\
& {\widetilde \Phi}^{(3)}(Z)=\int dZ_1 G(Z,Z_1;0){{\widetilde \Phi}^{(2)}}{}'(Z_1) \ .
\end{split}
\end{equation}
The $O(\epsilon^3)$ field strength is proportional to the pion mode solution \eqref{eq:solpion}. First note that
\begin{equation}
f_{Z\mu}^{(3)\,a}(Z,p_l)=\partial_Z A_\mu^{[2]\,(3)\,a }(Z,p_l)-i p_{l\,\mu} A_Z^{[2]\,(3)\,a}\propto \partial_Z \left( \Phi^{(3)}(Z) +q^2 {\widetilde \Phi}^{(3)}(Z)\right)-\frac{\Phi^{(2)}(Z)+q^2{\widetilde \Phi}^{(2)}(Z)}{1+Z^2}.
\end{equation}
On the other hand, using the definition of the Green's function in \eqref{eq:phi3}
\begin{equation}
\begin{split}
&\partial_Z \left[ u^3(Z) \partial_Z \Phi^{(3)}\right]={\Phi^{(2)}}'(Z) \\
&\partial_Z \left[ u^3(Z) \partial_Z {\widetilde \Phi}^{(3)}\right]={\widetilde \Phi}^{(2)}{}'(Z)\ .
\end{split}
\end{equation}
We can integrate once each equation and, since $u(Z)^3=1+Z^2=1/(\pi \phi_0(Z))$, it follows that
\begin{equation}
\begin{split}
& \partial_Z \Phi^{(3)}=\frac{\Phi^{(2)}(Z)}{1+Z^2}+c \pi \phi_0(Z)\ , \\
& \partial_Z {\widetilde \Phi}^{(3)}=\frac{{\widetilde \Phi}^{(2)}(Z)}{1+Z^2}+\tilde{c} \pi \phi_0(Z)\ .
\end{split}
\end{equation}
In the limit $Z\to \infty$ the terms proportional to $\Phi^{(2)}$, ${\widetilde \Phi}^{(2)}$ in the equations above are subleading, while the leading terms have the asymptotic form
\begin{equation}
\partial_Z\Phi^{(3)}\sim  c\pi \phi_0(Z) \sim \frac{c}{Z^2}\ , \ \partial_Z{\widetilde \Phi}^{(3)}\sim \tilde{c}\pi \phi_0(Z) \sim \frac{\tilde{c}}{Z^2}\ .
\end{equation}
Using the expansion in \eqref{eq:Gasymp}, the coefficients of the asymptotic terms are determined by the following integrals
\begin{equation}
\begin{split}
&c=\lim_{Z\to \infty} Z^2\partial_Z\Phi^{(3)}=\int_{-\infty}^\infty dZ_1\, \psi_0(Z_1) {\Phi^{(2)}}'(Z_1)= \frac{\pi}{12} \\
&\tilde{c}=\lim_{Z\to \infty} Z^2\partial_Z{\widetilde \Phi}^{(3)}= \int_{-\infty}^\infty dZ_1\, \psi_0(Z_1) {\widetilde \Phi}^{(2)}{}'(Z_1)\ .
\end{split}
\end{equation}
The first integral can easily be done taking into account that $\phi_0(Z)=\psi_0'(Z)$, so the integrand turns out to be a total derivative. The second integral can be manipulated to show it is equal to
\begin{equation}
\tilde{c}=\int_{-\infty}^\infty dZ_1\, \psi_0(Z_1) {\widetilde \Phi}^{(2)}{}'(Z_1)=-\int_{-\infty}^\infty d\tilde{Z} \frac{\left(\Phi^{(2)}(\tilde{Z})\right)^2}{u(\tilde{Z})}=-\frac{b}{(2\pi)^2} \ ,
\end{equation}
where $b$ is given in \eqref{eq:coefsSS}.

The leading pole contribution in the field strength can be identified as
\begin{equation}
 f_{Z\mu}^{(3)\,a}(Z,p_l)\simeq -\left[\frac{\pi^2}{12}-\frac{b}{4\pi} q^2\right]\phi_0(Z) \frac{p_{l\,\mu}p_l^\alpha}{p_l^2}i_\alpha^{(3)\,a}(p_i,p_j,p_k,p_l) \ .
\end{equation}
Plugging in \eqref{eq:i2mu} and \eqref{eq:i3mu}, and integrating over $q$ results in
\begin{equation}
 f_{Z\mu}^{(3)\,a}(Z,p_l)\simeq   \frac{4}{3}\phi_0(Z)\int_{p_i}\int_{p_j}\int_{p_k}\delta_{p_l-p_k-p_i-p_j}  \frac{p_{l\,\mu}p_i^\alpha p_j^\beta p_k^\gamma }{p_l^2 p_i^2 p_j^2 p_k^2}(p_l\cdot p_j) \left[ 1-\frac{12b}{\pi^3} (p_i\cdot p_j) \right] \epsilon^{abc}\epsilon^{bde} \hat{A}_{5\,\alpha}^d(p_i)\hat{A}_{5\,\beta}^e(p_j)\hat{A}_{5\,\gamma}^c(p_k) \ .
\end{equation}
Note that the radial dependence is the same as for the pion mode solution \eqref{eq:solpion}. Then, the calculation of the canonical momentum and expectation value of the current will proceed along similar steps, resulting in an exchange contribution to the axial current
\begin{equation}
 \vev{J_5^{\mu\,a}(p_l)}_e\simeq \frac{2}{3}\f^2\int_{p_i}\int_{p_j}\int_{p_k}\delta_{p_l-p_k-p_i-p_j}  \frac{p_l^\mu p_i^\alpha p_j^\beta p_k^\gamma }{p_l^2 p_i^2 p_j^2 p_k^2}(p_l\cdot p_j) \left[1-\frac{12 b}{\pi^3}  (p_i\cdot p_j) \right] \epsilon^{abc}\epsilon^{bde} \hat{A}_{5\,\alpha}^d(p_i)\hat{A}_{5\,\beta}^e(p_j)\hat{A}_{5\,\gamma}^c(p_k) \ .
\end{equation}
Restoring units, the exchange contribution to the leading pole in the four-point function of the current is
\begin{equation}
\begin{split}
&\vev{J_5^{\mu_1\,a_1}(p_1)J_5^{\mu_2\,a_2}(p_2)J_5^{\mu_3\,a_3}(p_3)J_5^{\mu_4\,a_4}(p_4)}_e  \\ 
 &\qquad\qquad\qquad \qquad\qquad\simeq   -2 i \f^2\left(\prod_{i=1}^4\frac{p_i^{\mu_i}}{p_i^2}\right) \delta_{\sum_{i=1}^4 p_i}\\
&\times  \Big[\Big\{(p_1\cdot p_2)-\frac{4b}{\pi^3 M_{KK}^2} \left[(p_1\cdot p_2)(p_2\cdot(p_3+p_4))-(p_1\cdot p_4)(p_2\cdot p_4)-(p_1\cdot p_3)(p_2\cdot p_3) \right] \Big\}\delta^{a_1 a_2}\delta^{a_3 a_4}\\
&+(2\leftrightarrow 3)+(2\leftrightarrow 4)\Big] \ .
\end{split}
 \end{equation}
Or, using momentum conservation
\begin{equation}
\begin{split}
&\vev{J_5^{\mu_1\,a_1}(p_1)J_5^{\mu_2\,a_2}(p_2)J_5^{\mu_3\,a_3}(p_3)J_5^{\mu_4\,a_4}(p_4)}_e  \\ 
 &\qquad\qquad\qquad \qquad\qquad\simeq   -2 i \f^2\left(\prod_{i=1}^4\frac{p_i^{\mu_i}}{p_i^2}\right) \delta_{\sum_{i=1}^4 p_i}\\
&\times  \Big[\Big\{(p_1\cdot p_2)-\frac{4b}{\pi^3} \left[(p_1\cdot p_3)^2+(p_1\cdot p_4)^2-2(p_1\cdot p_2)^2 \right] \Big\}\delta^{a_1 a_2}\delta^{a_3 a_4}\\
&+(2\leftrightarrow 3)+(2\leftrightarrow 4)\Big] \ .
\end{split}
 \end{equation}
 
%%%%%%%%%%%%%%%%%%%%%%%%%%%%%%%%%
\subsection{Contributions from $O(F^4)$ terms: vertex diagram}
%%%%%%%%%%%%%%%%%%%%%%%%%%%%%%%%%

The last possible contribution we have to study is originating from the $O(F^4)$ terms in the D$8$-brane action, the one that would introduce the non-linear terms in the equations
\begin{equation}
\begin{split}
 &I_Z^{[4]\,a}=\frac{1}{2}\left(\frac{\pi \alpha'}{L^2} \right)^2 u^2\left( \partial_\alpha \Pi^{[4]\,\alpha}_a+\frac{\delta {\cal L}_{\text{DBI}}^{[4]}}{\delta A_Z^a}\right) \\
 &I_\mu^{[4]\,a}=-\frac{1}{2}\left(\frac{\pi \alpha'}{L^2} \right)^2 \left[ \left(\partial_Z\left( u^2 \Pi^{[4]\,\mu}_a\right)-u^2\frac{\delta {\cal L}_{\text{DBI}}^{[4]}}{\delta A_\mu^a}\right)+ u^2\left( \partial_\alpha\left(\frac{\delta {\cal L}_{\text{DBI}}^{[4]} }{\delta \partial_\alpha A_\mu^a}\right) -\frac{\delta {\cal L}_{\text{DBI}}^{[4]}}{\delta A_\mu^a}\right)\right] \ .
\end{split}
\end{equation}
At $O(\epsilon^3)$ we need to keep only terms that are at most cubic in the fields, so terms $\sim \frac{\delta {\cal L}_{\text{DBI}}^{[4]}}{\delta A_M^a}$ can be dropped, and only terms involving three factors of the Abelianized field strengths $f_{MN}^{(1)\,a}$ remain. Among these, the leading pole contributions must come from terms with three factors of the $f_{Z\mu}^{(1)}$ components. One can check using \eqref{eq:DBIaction} that $\frac{\delta {\cal L}_{\text{DBI}}^{[4]} }{\delta \partial_\alpha A_\mu^a}$ does not introduce any such terms. Hence, using \eqref{eq:Pi4}, all leading pole contributions are the following
\begin{equation}
\begin{split}
 &I_Z^{[4]\,(3)\,a}\simeq -\frac{1}{6}\left(\frac{\pi \alpha'}{L^2} \right)^2 u^4 \eta^{\gamma\lambda}\eta^{\alpha\beta}\partial_\alpha\left(  f_{Z\gamma}^{(1)\,b} f^{(1)\,b}_{Z\lambda} f^{(1)\,a}_{Z\beta}+2f_{Z\gamma}^{(1)\,b} f^{(1)\,a}_{Z\lambda} f^{(1)\,b}_{Z\beta}\right) \\
 &I_\mu^{[4]\,(3)\,a}\simeq \frac{1}{6}\left(\frac{\pi \alpha'}{L^2} \right)^2\eta^{\gamma\lambda} \partial_Z\left( u^4\left(f_{Z\gamma}^{(1)\,b} f^{(1)\,b}_{Z\lambda} f^{(1)\,a}_{Z\mu}+2f_{Z\gamma}^{(1)\,b} f^{(1)\,a}_{Z\lambda} f^{(1)\,b}_{Z\mu} \right)\right) \ .
\end{split}
\end{equation}
Going to momentum space, and using \eqref{eq:fZ1} and \eqref{eq:tensionD8}, yields
\begin{equation}
\begin{split}
 &I_Z^{[4]\,(3)\,a}(Z,q)\simeq -u(Z)\phi_0(Z)^2 iq^\alpha j^{(3)\, a}_\alpha(p_i,p_j,p_k,q) \\
 &I_\mu^{[4]\,(3)\,a}(Z,q)\simeq  \partial_Z( u(Z)\phi_0(Z)^2 )  j^{(3)\, a}_\mu(p_i,p_j,p_k,q) \ ,
\end{split}
\end{equation}
where the leading pole factor is
\begin{equation}
\begin{split}\label{eq:j3mu}
 j^{(3)\, a}_\mu(p_i,p_j,p_k,q)\simeq &-\frac{3^5\pi}{4 \lambda_{\text{YM}}^2}\int_{p_i}\int_{p_j}\int_{p_k}\delta_{q-p_i-p_j-p_k} \frac{p_{k\,\mu} p_k^\nu p_i^\sigma p_j^\rho}{p_i^2p_j^2 p_k^2}(p_i\cdot p_j)\\
 &\times \left(\delta^{a_i a_j}\delta^{a a_k}+\delta^{a_i a_k}\delta^{a a_j} +\delta^{a_j a_k}\delta^{a a_i}\right)\hat{A}^{a_i}_{5\,\sigma}(p_i)\hat{A}^{a_j}_{5\,\rho}(p_j)\hat{A}^{a_k}_{5\,\nu}(p_k) \ .
\end{split}
\end{equation}
Following \eqref{eq:AparSol} and \eqref{eq:AperSol}, the $O(\epsilon^3)$ solution for the gauge potential is
\begin{equation}
\begin{split}
 &A_\mu^{[4]\,(3)\,a}(Z,p_l)\simeq \Psi^{(3)}(Z)\left(\delta_\mu^{\ \nu}-\frac{p_{l\,\mu}p_l^\alpha}{p_l^2} \right)j_\mu^{(3) \,a}(p_i,p_j,p_k,p_l)\\
 &A_Z^{[4]\,(3)\,a}(Z,p_l)=\frac{u(Z)\phi_0(Z)^2}{1+Z^2}\frac{i p_l^\alpha}{p_l^2} j_\alpha^{(3)\,a}(p_i,p_j,p_k,p_l) \ ,
\end{split}
\end{equation}
where
\bea
\Psi^{(3)}(Z) & = & \int dZ_1 G(Z,Z_1;0)\partial_{Z_1}\left( u(Z_1)\phi_0(Z_1)^2\right) \nonumber\\
  &= & \frac{7 Z \, _2F_1\left(\frac{1}{2},\frac{2}{3};\frac{3}{2};-Z^2\right)}{40 \pi ^2}+\frac{3Z \left(7 Z^2+11\right)}{40 \pi ^2 \left(Z^2+1\right)^{5/3}}-\frac{\Gamma \left(\frac{13}{6}\right) \arctan(Z)}{\pi ^{5/2} \Gamma \left(\frac{8}{3}\right)} \ .
\eea
The leading pole contribution in the field strength can be identified as
\begin{equation}
 f_{Z\mu}^{(3)\,a}(Z,p_l)\simeq -\frac{\Gamma\left(\frac{13}{6}\right)}{\pi^{3/2}\Gamma\left(\frac{8}{3}\right)}\phi_0(Z)\frac{p_{l\,\mu}p_l^\alpha}{p_l^2}j_\alpha^{(3)\,a}(p_i,p_j,p_k,p_l) \ .
\end{equation}
Plugging in \eqref{eq:j3mu} results in
\begin{equation}
\begin{split}
f_{Z\mu}^{(3)\,a}(Z,p_l)\simeq  & \phi_0(Z)\frac{3^5\Gamma\left(\frac{13}{6}\right)}{4\sqrt{\pi} \Gamma\left(\frac{8}{3}\right)\lambda_{\text{YM}}^2}\int_{p_i}\int_{p_j}\int_{p_k}\delta_{p_l-p_i-p_j-p_k} \frac{ p_{l\,\mu}p_k^\nu p_i^\sigma p_j^\rho }{p_i^2p_j^2 p_k^2p_l^2}(p_i\cdot p_j)(p_k\cdot p_l)\\
&\times \left(\delta^{a_i a_j}\delta^{a a_k}+\delta^{a_i a_k}\delta^{a a_j} +\delta^{a_j a_k}\delta^{a a_i}\right)\hat{A}^{a_i}_{5\,\sigma}(p_i)\hat{A}^{a_j}_{5\,\rho}(p_j)\hat{A}^{a_k}_{5\,\nu}(p_k) \ .
\end{split}
\end{equation}
Note that the radial dependence is, once more, the same as for the pion mode solution \eqref{eq:solpion}. Then, the calculation of the canonical momentum and expectation value of the current will proceed along similar steps, resulting in a vertex contribution to the axial current
\begin{equation}
\begin{split}
\vev{J_5^{\mu\,a}(p_l)}_v\simeq & \f^2\frac{3^5\Gamma\left(\frac{13}{6}\right)}{8\sqrt{\pi} \Gamma\left(\frac{8}{3}\right)\lambda_{\text{YM}}^2}\int_{p_i}\int_{p_j}\int_{p_k}\delta_{p_l-p_i-p_j-p_k} \frac{ p_l^\mu p_k^\nu p_i^\sigma p_j^\rho }{p_i^2p_j^2 p_k^2p_l^2}(p_i\cdot p_j)(p_k\cdot p_l)\\
&\times \left(\delta^{a_i a_j}\delta^{a a_k}+\delta^{a_i a_k}\delta^{a a_j} +\delta^{a_j a_k}\delta^{a a_i}\right)\hat{A}^{a_i}_{5\,\sigma}(p_i)\hat{A}^{a_j}_{5\,\rho}(p_j)\hat{A}^{a_k}_{5\,\nu}(p_k) \ .
\end{split}
\end{equation}
Restoring units, the vertex contribution to the leading pole in the four-point function of the current is
\begin{equation}
\begin{split}
&\vev{J_5^{\mu_1\,a_1}(p_1)J_5^{\mu_2\,a_2}(p_2)J_5^{\mu_3\,a_3}(p_3)J_5^{\mu_4\,a_4}(p_4)}_v\\ 
&\qquad\qquad\simeq   i \frac{\f^2}{M_{\text{KK}}^2}\frac{3^5\Gamma\left(\frac{13}{6}\right)}{4\sqrt{\pi} \Gamma\left(\frac{8}{3}\right)\lambda_{\text{YM}}^2} \left(\prod_{i=1}^4\frac{p_i^{\mu_i}}{p_i^2}\right) \delta_{\sum_{i=1}^4 p_i}\left(\delta^{a_1 a_2}\delta^{a_3 a_4}+\delta^{a_1 a_3}\delta^{a_2 a_4}+\delta^{a_1 a_4}\delta^{a_2 a_3}\right)\\ 
&\qquad\qquad\qquad\qquad \times \left[(p_1\cdot p_2)(p_3\cdot p_4)+(2\leftrightarrow 3)+(2\leftrightarrow 4) \right] \ .
\end{split}
\end{equation}

\bibliographystyle{JHEP}
\bibliography{refsscat}

\end{document}